\documentclass[]{aa}  

\usepackage{graphicx}
\usepackage{txfonts}
\usepackage{lscape}  
\usepackage{color}
\usepackage[colorlinks,citecolor=blue]{hyperref}

\usepackage{multirow}  

\begin{document}

\title{R-matrix electron-impact excitation data for the C-like iso-electronic sequence}


\author{Junjie Mao \inst{\ref{inst_strath}}
    \and N. R. Badnell \inst{\ref{inst_strath}}
    \and G. Del Zanna \inst{\ref{inst_damtp}}
    }

   \institute{Department of Physics, University of Strathclyde, Glasgow G4 0NG, UK\label{inst_strath}
   \and Department of Applied Mathematics and Theoretical Physics, University of Cambridge, Cambridge CB3 0WA, UK\label{inst_damtp}
   }


 
\abstract
{Emission and absorption features from C-like ions serve as temperature and density diagnostics of astrophysical plasmas. $R$-matrix electron-impact excitation data sets for C-like ions in the literature merely cover a few ions, and often only for the ground configuration.}
{Our goal is to obtain level-resolved effective collision strength over a wide temperature range for C-like ions from \ion{N}{II} to \ion{Kr}{XXXI} (i.e., N$^{+}$ to Kr$^{30+}$) with a systematic set of $R$-matrix calculations. We also aim to assess their accuracy. }
{For each ion, we included a total of 590 fine-structure levels in both the configuration interaction target and close-coupling collision expansion. These levels arise from 24 configurations $2l^3 nl^{\prime}$ with $n=2-4$, $l=0-1$, and $l^{\prime}=0-3$ plus the three configurations $2s^22p5l$ with $l=0-2$. The AUTOSTRUCTURE code was used to calculate the target structure. Additionally, the $R$-matrix intermediate coupling frame transformation method was used to calculate the collision strengths.}
{We compare the present results of selected ions with archival databases and results in the literature. The comparison covers energy levels, transition rates, and effective collision strengths. We illustrate the impact of using the present results on an \ion{Ar}{xiii} density diagnostic for the solar corona. The electron-impact excitation data is archived according to the Atomic Data and Analysis Structure (ADAS) data class adf04 and will be available in OPEN-ADAS. The data will be incorporated into spectral codes, such as CHIANTI and SPEX, for plasma diagnostics.}
{}
\titlerunning{Electron-impact excitation data for C-like ions}
\authorrunning{J.Mao et al.}       
\maketitle

\section{Introduction}
\label{sct:intro}
Emission and absorption features from C-like ions serve as temperature and density diagnostics for various types of astrophysical plasmas such as \citep[][]{mas84, mao17, dza18a}. Plasma models built on extensive and accurate atomic databases are essential to determine plasma parameters that span several orders of magnitudes in the parameter space. For instance, the density of photoionized outflows in the vinicity of black holes can vary from $\sim10^{3-5}~{\rm cm^{-3}}$ \citep[\ion{C}{III},][]{gab05, arv15} to $\gtrsim10^{6-14}~{\rm cm^{-3}}$ \citep[\ion{Si}{IX} and \ion{Fe}{XXI},][]{mil08,kin12,mao17} . Currently, the status of level-resolved electron-impact excitation data of C-like ions is rather poor. Such data are either lacking or obtained from distorted wave calculations in plasma codes, which are widely used in the community \citep[][]{mao19a}. 

More accurate $R$-matrix electron-impact excitation data for C-like ions are available in the literature, but only for a few ions and oftentimes only for the ground configuration. This is mainly because $R$-matrix calculations are rather computationally expensive. Therefore, $R$-matrix electron-impact excitation data for C-like ions are needed. 

\citet{gri98} introduced the $R$-matrix intermediate-coupling frame transformation (ICFT) method, which employs multi-channel quantum defect theory. The ICFT method first calculates the electron-impact excitation in pure $LS$-coupling, which subsequently, transforms into a relativistic coupling scheme via the algebraic transformation of the unphysical scattering or reactance matrices. Consequently, the ICFT method is inherently significantly faster than the classic Breit-Pauli $R$-matrix (BPRM) method \citep{ber95}, B-spline $R$-matrix (BSR) code \citep{zas06}, and Dirac atomic $R$-matrix code (DARC\footnote{http://amdpp.phys.strath.ac.uk/rmatrix/ser/darc/}). We refer readers to \citet{fme16}, \citet{agg17}, and \citet{dza19} for recent comparisons among different $R$-matrix methods and the impact on plasma diagnostics. 

In the past few decades, the $R$-matrix ICFT method has been used to perform large-scale calculations of electron-impact excitation data for a number of iso-electronic sequences: \citet[][Li-like]{lia11a}, \citet[][Be-like]{fme14a}, \citet[][B-like]{lia12}, \citet[][F-like]{wit07}, \citet[][Ne-like]{lia10}, \citet[][Na-like]{lia09}, and \citet[][Mg-like]{fme14b}. A review is presented by \citet{bad16}. We note that there are also other large-scale $R$-matrix calculations that cover individual ions in the C-like sequence, for instance,  \citep{lud10} and \citet{lia11b}. 

Here we present a systematic set of $R$-matrix ICFT calculations of C-like ions from \ion{N}{II} to \ion{Kr}{XXXI} (i.e., N$^{+}$ to Kr$^{30+}$) to obtain level-resolved effective collision strengths over a wide temperature range. Section~\ref{sct:mo} describes the atomic structure (Section~\ref{sct:str}) and collision calculations (Section~\ref{sct:col}). The results are summarized in Section~\ref{sct:res}. 
In Section~\ref{sct:dis}, we present comparisons between the present results for selected ions and some previous $R$-matrix calculations. This is followed by our summary in Section~\ref{sct:sum}. 

A supplementary package can be found at Zenodo \citep{mao19b}. This package includes the inputs of the AUTOSTRUCTURE and $R$-matrix ICFT calculations, atomic data from the present work, the archival database and literature, as well as scripts used to create the figures presented in this paper. 

\section{Method}
\label{sct:mo}
Following the previous case study of C-like \ion{Fe}{XXI} \citep{fme16}, for each ion, we include a total of 590 fine-structure levels (282 terms) in the configuration-interaction target expansion and close-coupling collision expansion. These levels (terms) arise from 27 configurations $2l^3 nl^{\prime}$ with $n=2-4$, $l=0-1$, and $l^{\prime}=0-3$ plus the 3 configurations $2s^22p5l$ with $l=0-2$ (Table~\ref{tbl:cfg}). 

\subsection{Structure}
\label{sct:str}
We use the AUTOSTRUCTURE code \citep{bad11} to calculate the target structure. The wave functions are calculated by diagonalizing the Breit-Pauli Hamiltonian \citep{eis74}. The one-body relativistic terms, mass-velocity, spin-orbit, and Darwin terms are included perturbatively. The Thomas-Fermi-Dirac-Amaldi model is used for the electronic potential. We adjust the $nl$-dependent scaling parameters \citep{nus78} in the following procedure without manual re-adjustment to avoid introducing arbitrary changes across the isoelectronic sequence. For each ion, we first optimize the scaling parameters of $1s$, $2s$, and $2p$ to minimize the equally-weighted sum of all LS term energies with $n=2$ (i.e., Conf.~1--3 in Table~\ref{tbl:cfg}. Since then, we fix the obtained scaling parameters of $1s$, $2s$, and $2p$. Subsequently, we optimize the scaling parameters of $3s$, $3p$ and $3d$ to minimize the equally-weighted sum of all LS term energies with $n=3$ (Conf.~4--12). We repeat this progressive procedure for $n=4$ (Conf.~13--24) and $n=5$ (15 configurations in total, including Conf.~25--27 in Table~\ref{tbl:cfg}). A similar optimization procedure was also used in \citet{lia11b} for instance. The scaling parameters of the 13 atomic orbitals (1s--5d) listed in Table~\ref{tbl:sca_par} are used for the structure (282 terms and 590 levels arising from 24 configurations) and the following collision calculation for all the ions ($Z=7-36$) in the sequence. 

Since the inner-region $R$-matrix codes require a unique set of non-relativistic orthogonal orbitals \citep{ber95}, we cannot exploit the full power of the general atomic structure codes. As shown later in Section~\ref{sct:dis}, the atomic structures obtained in the present work show relatively large differences with respect to experiment values, especially for the first few ions in the isoelectronic sequence, which require $R$-matrix calculations with pseudo-states. In general, this inaccuracy does not significantly affect plasma diagnostics using spectroscopically and astrophysically important lines \citep{dza19}.

\begin{table*}
\caption{List of configurations used for the collision calculations. }
\label{tbl:cfg}
\centering
\begin{tabular}{cl|cl|cl}
\hline\hline
\noalign{\smallskip} 
Index & Conf. & Index & Conf. & Index & Conf. \\
\noalign{\smallskip} 
\hline
\noalign{\smallskip} 
1 & $2s^22p^2$ & 2 & $2s2p^3$ & 3 & $2p^4$ \\
\noalign{\smallskip} 
4 & $2s^22p3s$ & 5 & $2s^22p3p$ & 6 & $2s^22p3d$ \\
\noalign{\smallskip} 
7 & $2s2p^23s$ & 8 & $2s2p^23p$ & 9 & $2s2p^23d$ \\
\noalign{\smallskip} 
10 & $2p^33s$ & 11 & $2p^33p$ & 12 & $2p^33d$ \\
\noalign{\smallskip} 
13 & $2s^22p4s$ & 14 & $2s^22p4p$ & 15 & $2s^22p4d$ \\
\noalign{\smallskip} 
16 & $2s^22p4f$ & 17 & $2s2p^24s$ & 18 & $2s2p^24p$ \\
\noalign{\smallskip} 
19 & $2s2p^24d$ & 20 & $2s2p^24f$ & 21 & $2p^34s$ \\
\noalign{\smallskip} 
22 & $2p^34p$ & 23 & $2p^34d$ & 24 & $2p^34f$ \\
\noalign{\smallskip} 
25 & $2s^22p5s$ & 26 & $2s^22p5p$ & 27 & $2s^22p5d$  \\
\noalign{\smallskip} 
\hline
\end{tabular}
\end{table*}

\begin{longtab}
\begin{landscape}
\begin{longtable}{lccccccccccccc}
\caption{\label{tbl:sca_par} Thomas-Fermi-Dirac-Amaldi potential scaling parameters used in the AUTOSTRUCTURE calculations for the C-like isoelectronic sequence. $Z$ is the atomic number, such as 8 for oxygen.}\\
\hline\hline 
\noalign{\smallskip} 
$Z$ & 1s & 2s & 2p & 3s & 3p & 3d & 4s & 4p & 4d & 4f & 5s & 5p & 5d \\ 
\noalign{\smallskip} 
\hline 
\noalign{\smallskip} 
 7 & 1.45393 & 1.23233 & 1.17669 & 1.26953 & 1.21267 & 1.25080 & 1.23008 & 1.18098 & 1.25066 & 1.34503 & 1.23349 & 1.16276 & 1.22628 \\ 
\noalign{\smallskip} 
 8 & 1.44061 & 1.23211 & 1.16447 & 1.25727 & 1.20910 & 1.28700 & 1.25818 & 1.13628 & 1.28178 & 1.45048 & 1.23717 & 1.19547 & 1.24916 \\ 
\noalign{\smallskip} 
 9 & 1.43087 & 1.21718 & 1.15800 & 1.26721 & 1.21498 & 1.28839 & 1.26055 & 1.20155 & 1.28081 & 1.41406 & 1.25850 & 1.26615 & 1.26167 \\ 
\noalign{\smallskip} 
10 & 1.42330 & 1.21797 & 1.15562 & 1.27452 & 1.20604 & 1.31551 & 1.25298 & 1.19051 & 1.28669 & 1.48826 & 1.22714 & 1.13335 & 1.27866 \\ 
\noalign{\smallskip} 
11 & 1.41849 & 1.21905 & 1.15436 & 1.28444 & 1.21802 & 1.31873 & 1.26513 & 1.24321 & 1.29576 & 1.30864 & 1.23342 & 1.17841 & 1.27008 \\ 
\noalign{\smallskip} 
12 & 1.41157 & 1.22000 & 1.15315 & 1.27551 & 1.22100 & 1.31086 & 1.22731 & 1.19577 & 1.30259 & 1.39045 & 1.26104 & 1.19319 & 1.27599 \\ 
\noalign{\smallskip} 
13 & 1.40714 & 1.22097 & 1.15331 & 1.27402 & 1.21814 & 1.31947 & 1.24394 & 1.19362 & 1.30126 & 1.00203 & 1.25662 & 1.20635 & 1.28309 \\ 
\noalign{\smallskip} 
14 & 1.40338 & 1.22185 & 1.15318 & 1.27151 & 1.21429 & 1.30879 & 1.23633 & 1.19412 & 1.29491 & 1.11223 & 1.25615 & 1.20015 & 1.27880 \\ 
\noalign{\smallskip} 
15 & 1.40008 & 1.22262 & 1.15319 & 1.27102 & 1.21451 & 1.30773 & 1.24854 & 1.20022 & 1.29219 & 1.18874 & 1.27157 & 1.20078 & 1.28423 \\ 
\noalign{\smallskip} 
16 & 1.39723 & 1.22336 & 1.15327 & 1.26997 & 1.21376 & 1.30684 & 1.25568 & 1.20533 & 1.29221 & 1.21814 & 1.26127 & 1.20173 & 1.28465 \\ 
\noalign{\smallskip} 
17 & 1.39479 & 1.22400 & 1.15340 & 1.26980 & 1.21339 & 1.30600 & 1.25006 & 1.21137 & 1.27265 & 1.24333 & 1.28481 & 1.21219 & 1.24540 \\ 
\noalign{\smallskip} 
18 & 1.39246 & 1.22459 & 1.15357 & 1.26950 & 1.21340 & 1.30751 & 1.27009 & 1.21436 & 1.30090 & 1.26759 & 1.23884 & 1.19126 & 1.28870 \\ 
\noalign{\smallskip} 
19 & 1.39059 & 1.22511 & 1.15374 & 1.26932 & 1.21345 & 1.30591 & 1.26970 & 1.21409 & 1.29127 & 1.28423 & 1.24265 & 1.20119 & 1.28844 \\ 
\noalign{\smallskip} 
20 & 1.38889 & 1.22559 & 1.15394 & 1.26920 & 1.21354 & 1.30553 & 1.26773 & 1.21324 & 1.29214 & 1.29210 & 1.24721 & 1.20131 & 1.28923 \\ 
\noalign{\smallskip} 
21 & 1.38727 & 1.22602 & 1.15413 & 1.26910 & 1.21365 & 1.30518 & 1.32079 & 1.28093 & 1.39270 & 1.32000 & 1.22758 & 1.20778 & 1.29012 \\ 
\noalign{\smallskip} 
22 & 1.38584 & 1.22642 & 1.15433 & 1.26903 & 1.21376 & 1.30487 & 1.26575 & 1.21595 & 1.29368 & 1.30745 & 1.23947 & 1.20646 & 1.29044 \\ 
\noalign{\smallskip} 
23 & 1.38453 & 1.22679 & 1.15451 & 1.26898 & 1.21388 & 1.30458 & 1.26578 & 1.21543 & 1.29269 & 1.31435 & 1.24636 & 1.20781 & 1.29041 \\ 
\noalign{\smallskip} 
24 & 1.38343 & 1.22712 & 1.15470 & 1.26894 & 1.21400 & 1.30431 & 1.26553 & 1.21571 & 1.29327 & 1.31872 & 1.25001 & 1.21009 & 1.29077 \\ 
\noalign{\smallskip} 
25 & 1.38240 & 1.22744 & 1.15487 & 1.26891 & 1.21411 & 1.30407 & 1.26571 & 1.21598 & 1.29308 & 1.32287 & 1.25213 & 1.21166 & 1.29093 \\ 
\noalign{\smallskip} 
26 & 1.38137 & 1.22772 & 1.15505 & 1.26889 & 1.21422 & 1.30384 & 1.26594 & 1.21773 & 1.29318 & 1.32517 & 1.25400 & 1.21291 & 1.29126 \\ 
\noalign{\smallskip} 
27 & 1.38038 & 1.22799 & 1.15518 & 1.26888 & 1.21433 & 1.30363 & 1.26666 & 1.21738 & 1.29304 & 1.32910 & 1.25520 & 1.21392 & 1.29173 \\ 
\noalign{\smallskip} 
28 & 1.37956 & 1.22824 & 1.15534 & 1.26887 & 1.21444 & 1.30343 & 1.26602 & 1.21748 & 1.29340 & 1.33299 & 1.25636 & 1.21483 & 1.29158 \\ 
\noalign{\smallskip} 
29 & 1.37877 & 1.22847 & 1.15550 & 1.26886 & 1.21454 & 1.30325 & 1.26611 & 1.21749 & 1.29338 & 1.33353 & 1.25684 & 1.21423 & 1.29197 \\ 
\noalign{\smallskip} 
30 & 1.37803 & 1.22869 & 1.15564 & 1.26886 & 1.21464 & 1.30308 & 1.26649 & 1.21778 & 1.29434 & 1.33555 & 1.25794 & 1.21616 & 1.29264 \\ 
\noalign{\smallskip} 
31 & 1.37736 & 1.22889 & 1.15579 & 1.26886 & 1.21473 & 1.30292 & 1.26664 & 1.21790 & 1.29358 & 1.33756 & 1.25863 & 1.21672 & 1.29207 \\ 
\noalign{\smallskip} 
32 & 1.37672 & 1.22908 & 1.15592 & 1.26886 & 1.21482 & 1.30277 & 1.26680 & 1.21816 & 1.29357 & 1.33930 & 1.25925 & 1.21696 & 1.29369 \\ 
\noalign{\smallskip} 
33 & 1.37613 & 1.22926 & 1.15605 & 1.26886 & 1.21491 & 1.30263 & 1.26695 & 1.21842 & 1.29359 & 1.34089 & 1.25981 & 1.21749 & 1.29423 \\ 
\noalign{\smallskip} 
34 & 1.37557 & 1.22943 & 1.15618 & 1.26887 & 1.21499 & 1.30250 & 1.26710 & 1.21866 & 1.29361 & 1.34235 & 1.26032 & 1.21798 & 1.29392 \\ 
\noalign{\smallskip} 
35 & 1.37504 & 1.22959 & 1.15630 & 1.26888 & 1.21507 & 1.30237 & 1.26718 & 1.21889 & 1.29363 & 1.34368 & 1.26078 & 1.21844 & 1.29436 \\ 
\noalign{\smallskip} 
36 & 1.37454 & 1.22974 & 1.15641 & 1.26888 & 1.21515 & 1.30226 & 1.26728 & 1.21910 & 1.29365 & 1.34477 & 1.26120 & 1.21885 & 1.29444 \\ 
\noalign{\smallskip} 
\hline 
\end{longtable}
\end{landscape}
\end{longtab}

\subsection{Collision}
\label{sct:col}
The $R$-matrix collision calculation consists of the inner and outer-region calculations \citep{bur11}. The inner-region calculation is further split into the exchange and non-exchange calculations. Following the previous case study of C-like \ion{Fe}{XXI} \citep{fme16}, we include angular momenta up to $2J=23$ and $2J=77$ for the inner-region exchange and non-exchange calculation, respectively, for the entire isoelectronic sequence. For higher angular momenta up to infinity, we use the top-up formula of the Burgess sum rule \citep{bur74} for dipole allowed transitions, and a geometric series for the remaining non-forbidden (i.e., non-dipole allowed) transitions \citep{bad01a}. 

The outer-region calculation is split into a fine energy mesh exchange calculation, a coarse energy mesh exchange calculation, and a (coarse energy mesh) non-exchange calculation. A fine energy mesh is used between the first and last thresholds for the outer-region exchange calculation to sample the resonances. With an increasing ionic charge, the number of sampling points in the fine energy mesh increases from $\sim3600$ for \ion{N}{II} to $\sim30000$ for \ion{Kr}{XXXI} to strike the balance between the computational cost and resonance sampling. Along the iso-electronic sequence, in the resonance region, the characteristic scattering energy increases by a factor of $z^2$, with $z$ the ionic charge (such as $z=3$ for \ion{O}{III} and $z=20$ for \ion{Fe}{XXI}). However, the autoionization width of the resonance remains constant. That is to say, to resolve the resonance region to the same degree, the step size of the energy mesh needs to be reduced by a factor of $z^2$ with increasing $z$. To avoid unreasonable computation cost of high-$z$ ions, following \citet{wit07}, we reduce the step size of the energy mesh by a factor of $z$ (see also Appendix~\ref{sct:eng_mesh}).

A coarse energy mesh, with $\sim1000$ points for all the ions along the iso-electronic sequence is used from the last threshold up to $\sim3I_p$, where $I_p$ is the ionization potential in units of Rydberg This allows us to determine a smooth background of the outer-region exchange calculation. 

Another coarse energy mesh with $\sim1400$ points for all the ions along the iso-electronic sequence is used from the first threshold up to $\sim3I_p$ for the outer-region non-exchange calculation. Since this coarse energy mesh covers the resonance region, it is possible that unresolved resonance(s) appear in the ordinary collision strength of the outer-region non-exchange calculation. Therefore, post-processing to remove the unresolved resonance(s) is necessary.  

The effective collision strength ($\Upsilon_{ij}$) for electron-impact excitation is obtained by convolving the ordinary collision strength ($\Omega_{ij}$) with the Maxwellian velocity distribution:
\begin{equation}
    \Upsilon_{ij} = \int \Omega_{ij}~\exp\left(-\frac{E}{kT}\right)~d\left(\frac{E}{kT}\right)~,
\label{eq:upsilon}
\end{equation}
where $E$ is the kinetic energy of the scattered free electron, $k$ the Boltzmann constant, and $T$ the electron temperature of the plasma. 

To obtain effective collision strengths at high temperatures, ordinary collision strengths at high collision energies are required, which is inefficient to be calculated with the $R$-matrix method. Hence, we use AUTOSTRUCTURE to calculate the infinite-energy Born and radiative dipole limits. Between the last calculated energy point and the two limits, we interpolate taking into account the type of transition in the Burgess-Tully scaled domain \citep[i.e., the quadrature of reduced collision strength over reduced energy,][]{bur92} to complete the Maxwellian convolution (Equation~\ref{eq:upsilon}).

\section{Results}
\label{sct:res}
We obtain $R$-matrix electron-impact excitation data for the C-like iso-electronic sequence from \ion{N}{II} to \ion{Kr}{XXXI} (i.e., ${\rm N^+}$ and ${\rm Kr^{30+}}$). Our effective collision strengths cover a wide range of temperature $(z+1)^2(2\times10^1,~2\times10^6)~{\rm K}$. They are to be applied to astrophysical plasmas in various conditions.

The ordinary collision strengths will be archived in OPEN-ADAS \footnote{http://open.adas.ac.uk/}. The effective collision strengths are archived according to the Atomic Data and Analysis Structure (ADAS) data class adf04 and will be available in OPEN-ADAS
and our UK-APAP website\footnote{http://apap-network.org/}. These data will be incorporated into plasma codes like CHIANTI \citep{der97,der19} and SPEX \citep{kaa96,kaa18} for plasma diagnostics.

\section{Discussion}
\label{sct:dis}
We selected four ions \ion{O}{III}, \ion{Ne}{V}, \ion{Si}{XI}, and \ion{Fe}{XXI} across the iso-electronic sequence to illustrate the quality of our structure and collision calculation. These ions were selected because detailed results from archival databases (NIST\footnote{https://www.nist.gov/pml/atomic-spectra-database}, MCHF\footnote{https://nlte.nist.gov/MCHF/}, and OPEN-ADAS) and the literature are available for comparison purposes.

\begin{figure*}
\centering
\includegraphics[width=.7\hsize, trim={1.cm 0.5cm 1.5cm 0.65cm}, clip, angle=90]{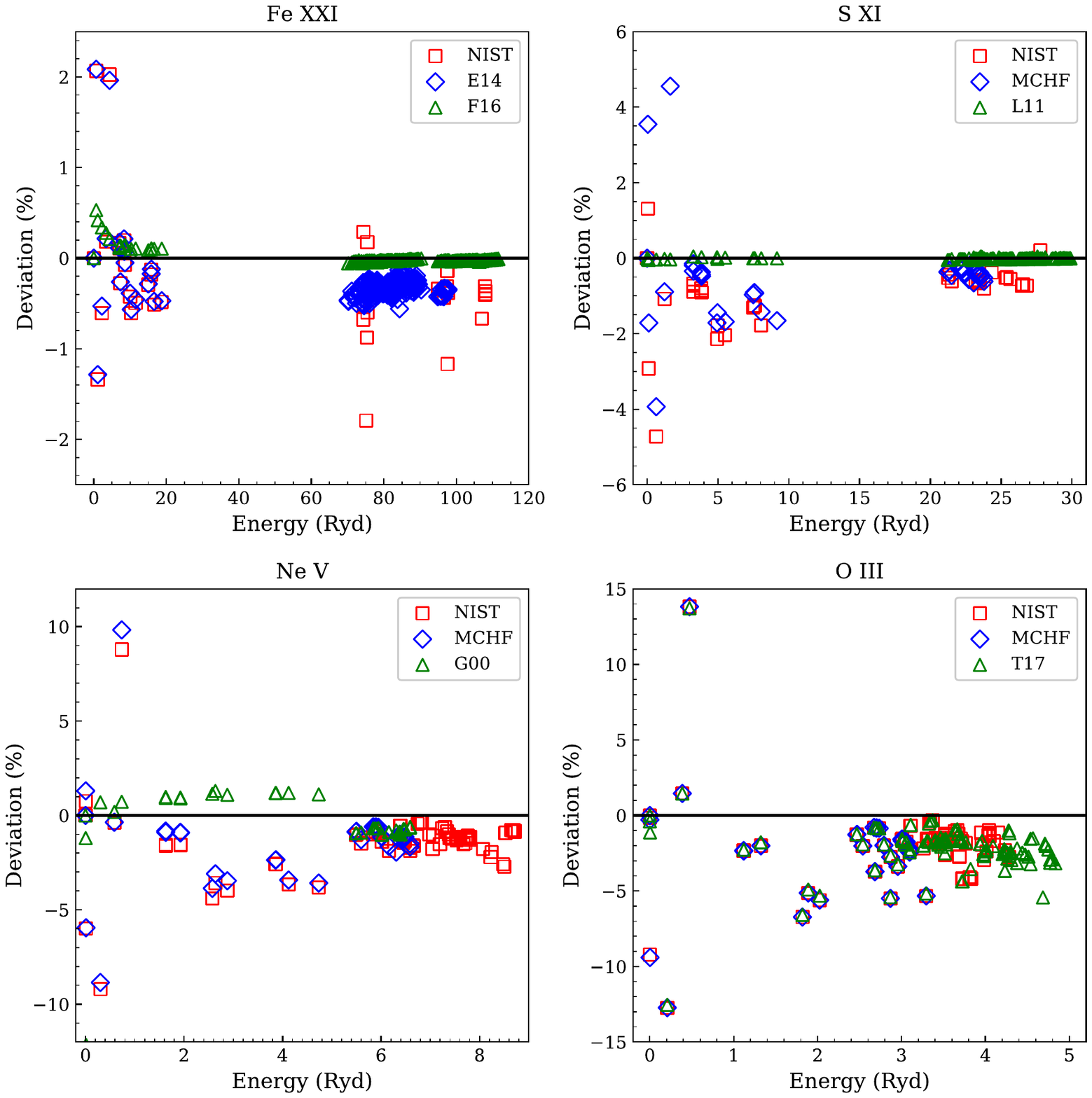}
\caption{Percentage deviations between the present energy levels (horizontal lines in black), the experimental ones (NIST) and other theoretical values as available in archival databases (MCHF, OPEN-ADAS) and previous works: F16 refers to \citet{fme16}, E14 refers to \citet{ekm14}, L11 refers to \citet{lia11b}, G00 refers to \citet[][OPEN-ADAS]{gri00}, and T17 refers to \citet{tay17}. }
\label{fig:sca_lev_clike}
\end{figure*}

For each ion (Sections~\ref{sct:062621} to \ref{sct:060803}), we first compare the energy levels. Figure~\ref{fig:sca_lev_clike} illustrates the deviation (in percent) of the energy levels in archival databases and previous works with respect to the present work. A histogram plot of the data shown in Figure~\ref{fig:sca_lev_clike} is also shown in Appendix~\ref{sct:hist_str}. Generally speaking, the energy levels of the present work agree well ($\lesssim5\%$) with archival databases and previous works for high-charge ions. A larger deviation ($\lesssim15\%$) is found for low-charge ions, in particular, for some of the lowest lying energy levels.

Transition strengths are also compared. The oscillator strength ($f_{ij}$), which is related to the A-value (i.e., the Einstein coefficient), is often used, 
\begin{equation}
    f_{ij} = \frac{mc}{8 \pi^2 e^2}\lambda_{ij}^2 \frac{g_j}{g_i} A_{ji},
    \label{eq:osc}
\end{equation}
where $m$ and $e$ are the rest mass and charge of electron, respectively, $c$ the speed of light, $g_j$ and $g_i$ the statistical weights of the upper ($j$) and lower ($i$) levels, respectively, and $\lambda_{ij}$ the wavelength of the transition $i-j$. 

\begin{figure*}
\centering
\includegraphics[width=.7\hsize, trim={1.cm 0.5cm 1.5cm 0.65cm}, clip, angle=90]{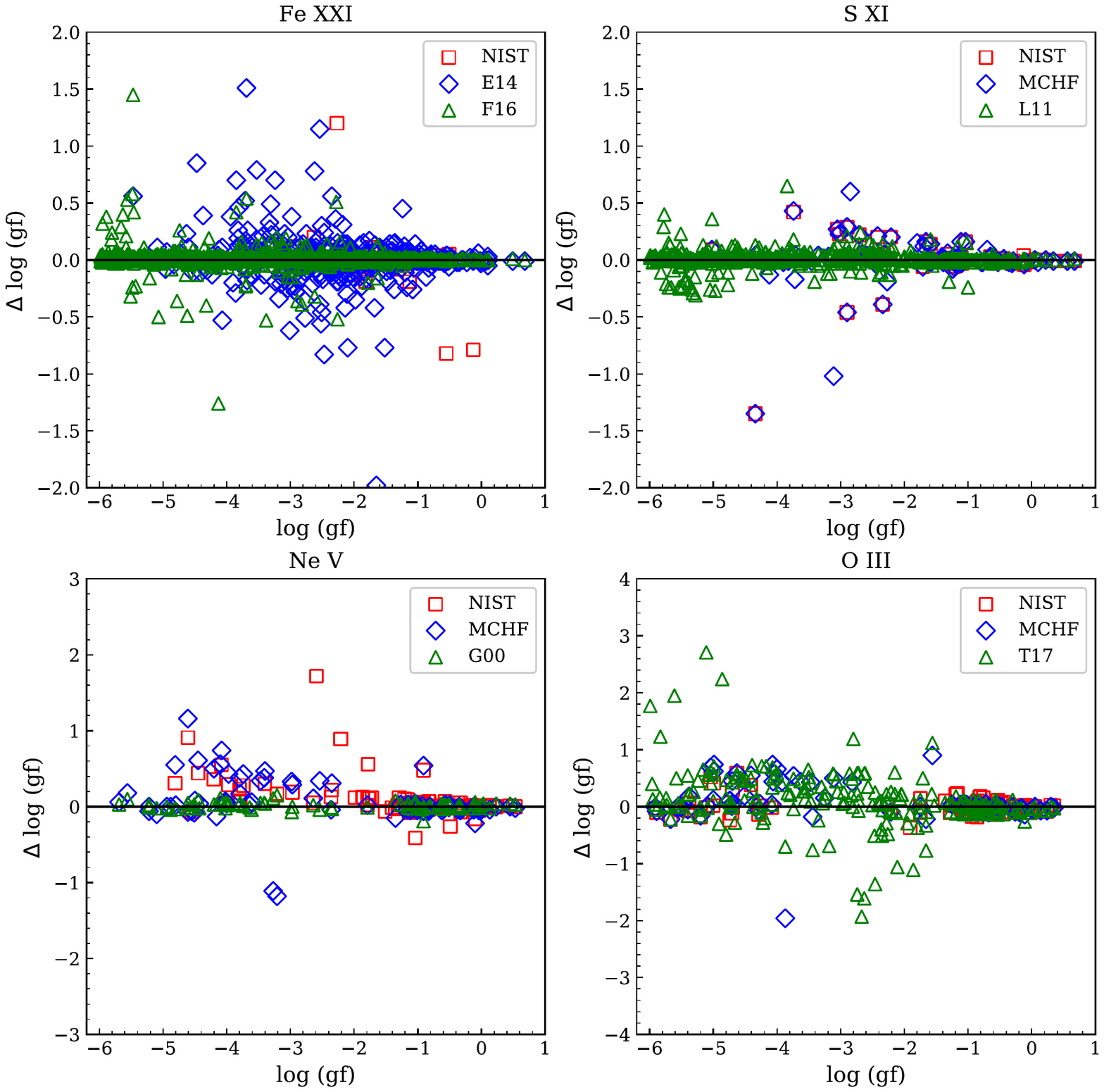}
\caption{Comparisons of $\log~(gf)$ from the present work (black horizontal line) with archival databases and previous works. F16 refers to \citet{fme16}, E14 refers to \citet{ekm14}, L11 refers to \citet{lia11b}, G00 refers to \citet[][OPEN-ADAS]{gri00}, and T17 refers to \citet{tay17}. We note that this comparison is limited to relatively strong transitions with $\log~(gf) \gtrsim 10^{-6}$ originating from the lowest five energy levels.}
\label{fig:sca_tran_clike}
\end{figure*}

Figure~\ref{fig:sca_tran_clike} shows the deviation $\Delta \log~(gf)$ of archival databases and previous studies with respect to the present work. A histogram plot of the data shown in Figure~\ref{fig:sca_tran_clike} is shown in Appendix~\ref{sct:hist_str}. We limit the comparison to relatively strong transitions with $\log~(gf) \gtrsim 10^{-6}$ from the lowest five energy levels of the ground configuration: $2s^22p^2 ~(^3P_{0-2},^1D_2,^1S_0)$ as they are metastable levels. For those weak transitions excluded in our comparison, $\log~(gf)$ might differ by several orders of magnitude among archival databases, previous studies, and the present work. This is often due to the different number of configuration interaction levels included, as well as the method adopted. Nonetheless, the weak transitions are not expected to significantly impact the plasma diagnostics as the five metastable levels drive the population of all the other levels in the C-like ions, for astrophysical plasma.

Subsequently, we compare the collision data for \ion{Fe}{XXI} (Section~\ref{sct:062621}), \ion{S}{XI} (Section~\ref{sct:061611}), \ion{Ne}{V} (Section~\ref{sct:061005}), and \ion{O}{III} (Section~\ref{sct:060803}) with previous $R$-matrix calculations. Hexbin plots\footnote{To represent the relationship of two large sets of numerical variables, instead of overlapping data points in a scatter plot, the hexbin plotting window is split into hexbins, and the number of points per hexbin is counted and color coded. The supplementary package on Zenodo \citep{mao19b} provides scripts to reproduce the hexbin plots presented in this paper. A simple demo of hexbin plot is also available \href{https://matplotlib.org/3.1.0/gallery/statistics/hexbin_demo.html}{here}. } \citep{car87} are used to compare the effective collision strengths of a large number of transitions. In Section~\ref{sct:061813}, we compare the collision data for \ion{Ar}{XIII} with a previous distorted wave calculation. Finally, we demonstrate the impact on the density diagnostics using these two data sets of \ion{Ar}{XIII}. 

We note that all $R$-matrix calculations (including the present calculation) without pseudo-states are not converged for the high-lying levels, both with respect to the $N$-electron target configuration interaction expansion and the ($N+1$)-electron close-coupling expansion. Here we include configurations up to $n=4$ (24 in total) in addition to three configurations with $n=5$ (Table~\ref{tbl:cfg}). Accordingly, the present effective collision strengths involving energy levels with $n\le3$ are better converged than those with $n\ge4$. Future larger-scale $R$-matrix ICFT calculations or $R$-matrix calculations with pseudo-state calculations can improve the accuracy of transitions involving the high-lying levels, especially amongst the high-lying levels.

\subsection{\ion{Fe}{XXI}}
\label{sct:062621}
\begin{figure*}
\centering
\includegraphics[width=.7\hsize, trim={0.5cm 0.5cm 0.5cm 0.62cm}, clip]{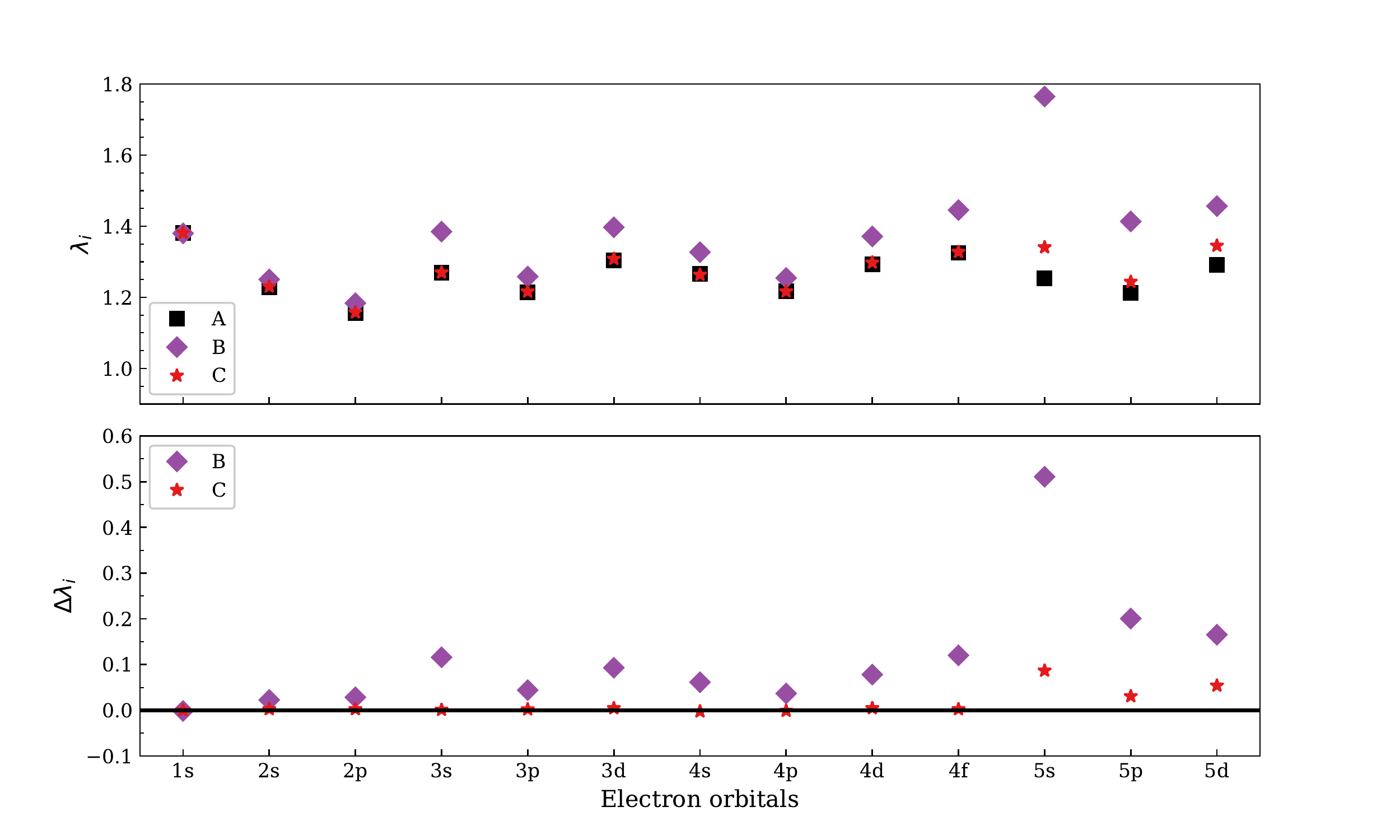}
\caption{Three sets of scaling parameters for \ion{Fe}{XXI}. The black squares (set A) correspond to those listed in Table~\ref{tbl:sca_par}. The purple diamonds (set B) correspond to the scaling parameters used in \citet{fme16}. The red stars (set C, see Section~\ref{sct:062621}) correspond to the third set of scaling parameters which have a smaller deviation with respect to the default set. The level energies, A-values, and effective collision strengths shown in Table~\ref{tbl:062621_78and79} are very sensitive to the scaling parameters of $3s$ and $3d$. }
\label{fig:cf_lambda_062621}
\end{figure*}

The most recent calculation of $R$-matrix electron-impact excitation data for \ion{Fe}{XXI} (or ${\rm Fe^{20+}}$) is presented by \citet[][F16 hereafter]{fme16}, including 590 fine-structure levels in both the configuration interaction and close-coupling expansions. We limit our comparison to F16 and refer readers to F16 for their comparison with other previous calculations \citep{agg99a, but00, bad01b}. 
Both F16 and the present work use the AUTOSTRUCTURE code for the structure calculation. Although both calculations include 590 fine-structure levels in the configuration interaction and close-coupling expansions, different scaling parameters (data sets A and B in Figure~\ref{fig:cf_lambda_062621}) lead to slightly different atomic structures. As shown in the top-left panel of Figure~\ref{fig:sca_lev_clike}, generally speaking, the energy levels of the present work and F16 agree within $\lesssim0.1\%$. The first few levels have a slightly larger deviation of $\lesssim0.5\%$, yet smaller than the $\lesssim2\%$ deviation with respect to NIST and \cite[][E14 hereafter]{ekm14}. Additionally, there is a shift of $\sim0.3$~Ryd between E14 and the present work (and F16) for the $2s^22p3s~(3P_0)$.

Similarly, as shown in the top-left panel of Figure~\ref{fig:sca_tran_clike}, transition strengths agree well among NIST, F16, and the present work with merely a few exceptions. Larger deviations are found between E14 and the present work.

\begin{figure*}
\centering
\includegraphics[width=.6\hsize, trim={4.5cm 1.0cm 1.0cm 1.0cm}, clip, angle=90]{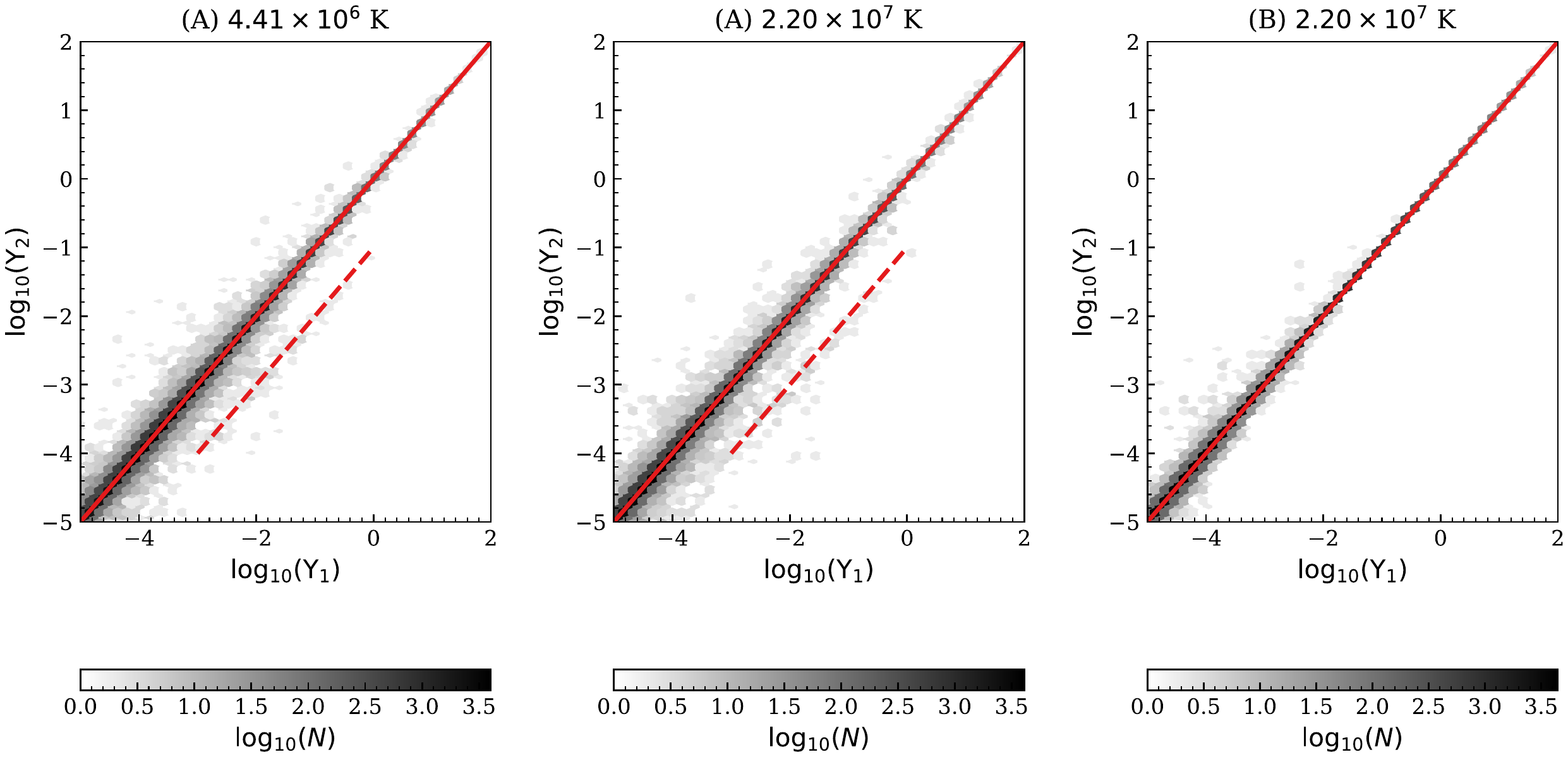}
\caption{Hexbin plots of the comparison of the \ion{Fe}{XXI} (or ${\rm Fe^{20+}}$) effective collision strengths between two sets (A and B) of calculations ($\Upsilon_1$) and \citet[][$\Upsilon_2$]{fme16} at relatively high temperatures. Data set A is the default of the present work. In data set B, we use the scaling parameters of \citet{fme16} (Figure~\ref{fig:cf_lambda_062621}). The darker the color, the greater number of transitions $\log_{10}(N)$. The diagonal line in red indicates $\Upsilon_1=\Upsilon_2$. The dashed lines in red highlight transitions in the ``long island" (see Section~\ref{sct:062621} for more details). }
\label{fig:hb_ecs_062621_mao19_fme16_high}
\end{figure*}

The scattering calculations of both the present work and F16 were performed with the $R$-matrix ICFT method. We remind the readers that the atomic structures are slightly different between the two calculations. Figure~\ref{fig:hb_ecs_062621_mao19_fme16_high} shows the comparison of the effective collision strengths at relatively high temperatures ($4.41\times10^6~{\rm K}$ and $2.20\times10^7~{\rm K}$). There are in total $\sim1.3-1.4\times10^5$ transitions with $\log (gf) > -5$ in both data sets for all three panels. In the left and middle panels, $\sim5-8\%$, $\sim1\%$, and $\sim0.1\%$ have deviation larger than 0.1 dex, 0.3 dex, and 1 dex, respectively. In the right panel, $\sim3\%$, $\sim0.2\%$, and $0.01\%$ have deviation larger than 0.1 dex, 0.3 dex, and 1 dex, respectively.

In the left and middle panels of Figure~\ref{fig:hb_ecs_062621_mao19_fme16_high}, the ``long island" in parallel to yet below the diagonal line in red is mainly contributed by transitions involving level \#78 ($2s2p^23s,~^3P_1$) and \#79 ($2s2p^23d,~^5D_1$). When we use the scaling parameters of F16 (set B) the long island is no longer present (the right panel of Figure~\ref{fig:hb_ecs_062621_mao19_fme16_high}). We performed another calculation (set C) with a third set of scaling parameters\footnote{These scaling parameters were determined by following the progressive optimization procedure described in Section~\ref{sct:str} but for configurations with $n=5$, we only include the lowest three (instead of 15 in total) configurations.}, which has a smaller deviation with respect to the default calculation (set A). This shows that the level energies, A-values, and effective collision strength in the ``long island" are very sensitive to the scaling parameters of $3s$ and $3d$ (Table~\ref{tbl:062621_78and79}). When we compare the energies and A-values with respect to CHIANTI and SPEX (Table~\ref{tbl:062621_78and79}), our default calculation (A) are comparable in terms of energies, and slightly ``better" than set B yet slightly ``worse" than set C in terms of A-values. 

\begin{table*}
\caption{Energy (${\rm cm^{-1}}$), $A$-value (${\rm s^{-1}}$), $\Upsilon$ (at $4.41\times10^6~{\rm K}$), and $\Upsilon$ (at $\infty$) for levels $2s~2p^2~3s,~^3P_1$ (\#78 in the present work) and $2s~2p^2~3d,~^5D_1$ (\#79) of \ion{Fe}{XXI}. The scaling parameters of $3s$ and $3d$, most relevant to the ``long island" shown in Figure~\ref{fig:hb_ecs_062621_mao19_fme16_high}, are also tabulated. }
\label{tbl:062621_78and79}
\large
\centering
\begin{tabular}{cccccc}
\noalign{\smallskip}
\hline\hline
\noalign{\smallskip}
ID & $\lambda~(3s)$ & Energy (\#78) & A(21-78) & $\Upsilon(21-78)$ & $\Upsilon(21-78)$  \\
    &   & ${\rm cm^{-1}}$ & ${\rm s^{-1}}$ &  @$4.41\times10^6~{\rm K}$ & @$\infty$  \\
\noalign{\smallskip}
\hline
\noalign{\smallskip}
A & 1.26889 & 8577023 & $9.17\times10^{9}$ & $9.67\times10^{-2}$ & $2.67\times10^{-2}$ \\  
B & 1.38480 & 8573268 & $1.47\times10^{10}$ & $1.54\times10^{-1}$ & $4.28\times10^{-2}$ \\ 
C & 1.26986 & 8576789 & $7.69\times10^{9}$ & $8.12\times10^{-2}$ & $2.24\times10^{-2}$ \\ 
\noalign{\smallskip}
CHIANTI & -- -- & 8664021 (\#97) & $2.59\times10^9$ & -- -- & -- -- \\ 
SPEX & -- -- & 8547294 (\#80) & $1.36\times10^9$ & -- -- & -- -- \\ 
\noalign{\smallskip}
\hline
\noalign{\smallskip}
ID & $\lambda~(3d)$ & Energy (\#79) & A(21-79) & $\Upsilon(21-79)$ & $\Upsilon(21-79)$  \\
\noalign{\smallskip}
\hline
\noalign{\smallskip}
A & 1.30384 & 8577328 & $5.92\times10^{9}$ & $6.18\times10^{-2}$ & $1.72\times10^{-2}$ \\ 
B & 1.39690 & 8574110 & $4.30\times10^{8}$ & $4.50\times10^{-3}$ & $1.25\times10^{-3}$ \\ 
C & 1.30830 & 8577085 & $7.40\times10^{9}$ & $7.73\times10^{-2}$ & $2.15\times10^{-2}$ \\ 
\noalign{\smallskip}
CHIANTI & -- -- & 8664020 (\#78) & $2.59\times10^9$ & -- -- & -- -- \\ 
SPEX & -- -- & 8547294 (\#78) & $1.36\times10^9$ & -- -- & -- -- \\ 
\noalign{\smallskip}
\hline
\end{tabular}
\tablefoot{Data set A is the default of the present work. In data set B, we use the scaling parameters of \citet{fme16}. In data set C, we use a third set of scaling parameters (see Figure~\ref{fig:cf_lambda_062621}). For CHIANTI, database v9.0 \citep{der19} is used and 620 energy levels are included for \ion{Fe}{XXI}. Theoretical energy levels \#78 and \#97 are collected from \citet{lan06}. For SPEX, database SPEXACT v3.05 \citep{kaa18} is used and and 1400 energy levels are included for \ion{Fe}{XXI}. No information of levels \#21 $2s^2~2p~3s,~(^3P_0)$, $2s~2p^2~3s,~(^3P_1)$, and $2s~2p^2~3d,~(^5D_1)$ are available in NIST v5.6.1 \citep{kra18}. }
\end{table*}

\begin{figure*}
\centering
\includegraphics[width=.6\hsize, trim={4.5cm 1.0cm 1.0cm 1.0cm}, clip, angle=90]{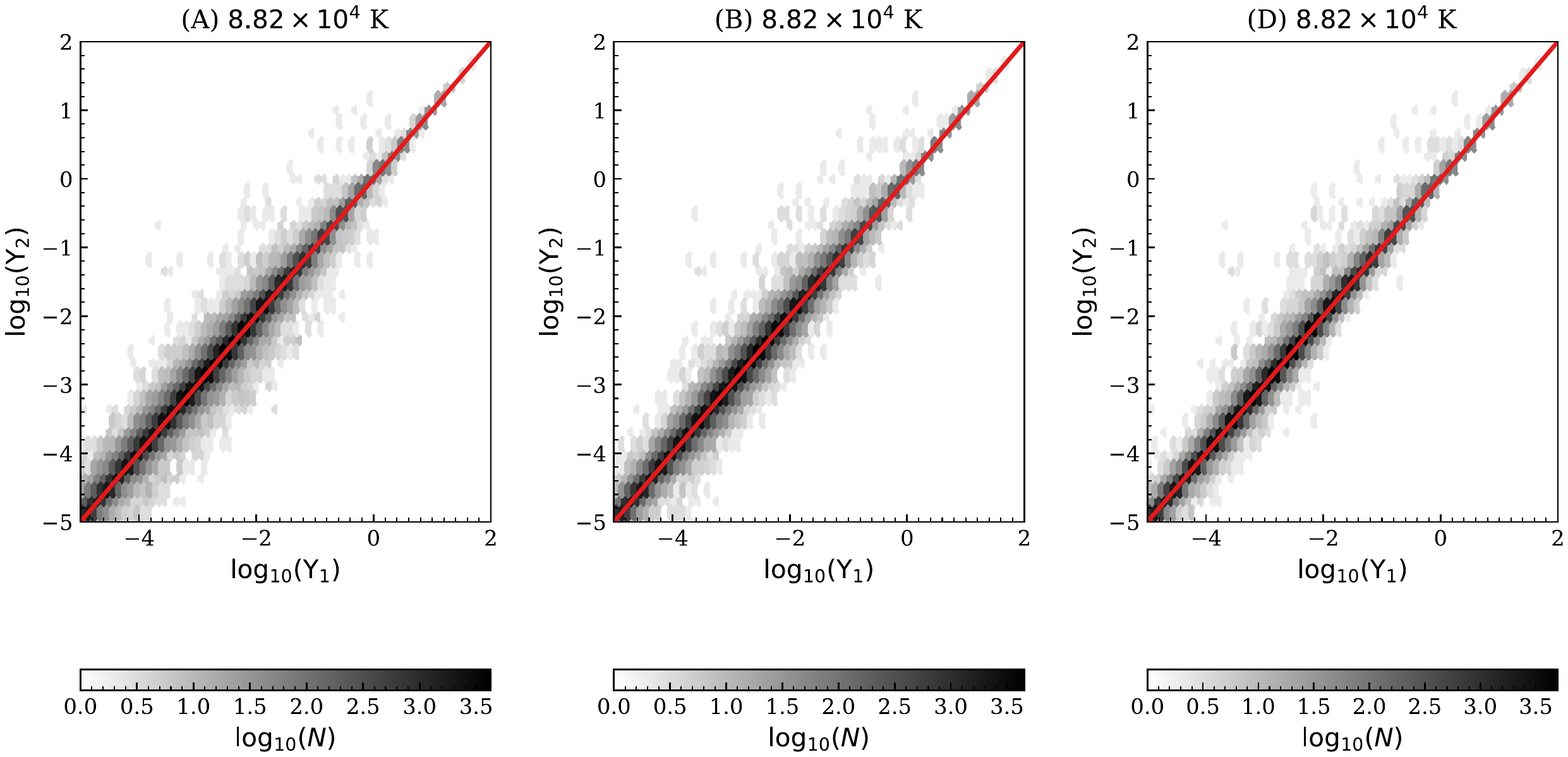}
\caption{Hexbin plots of the comparison of the \ion{Fe}{XXI} (or ${\rm Fe^{20+}}$) effective collision strengths between three sets (A, B, and D) of present calculations ($\Upsilon_1$) and \citet[][F16, $\Upsilon_2$]{fme16} at a relatively low temperature. Data set A is the default of the present work. In data set B, we use the scaling parameters of \citet{fme16} (Figure~\ref{fig:cf_lambda_062621}). In data set D, we use the same scaling parameters and the same number of points (four times our default calculation) for the fine energy mesh, as in F16. The darker the color, the greater number of transitions $\log_{10}(N)$. The diagonal line in red indicates $\Upsilon_1=\Upsilon_2$}. 
\label{fig:hb_ecs_062621_mao19_fme16_low}
\end{figure*}

At relatively low temperatures, a large deviation is still observed even when we use the scaling parameters of F16 (the middle panel of Figure~\ref{fig:hb_ecs_062621_mao19_fme16_low}. There are in total $\sim1.6\times10^5$ transitions with $\log (gf) > -5$ in both data sets for all three panels. In the left panel, $\sim20\%$, $\sim3\%$, and $\sim0.1\%$ have deviation larger than 0.1 dex, 0.3 dex, and 1 dex, respectively. In the middle panel, $\sim20\%$, $\sim2\%$, and $\sim0.06\%$ have deviation larger than 0.1 dex, 0.3 dex, and 1 dex, respectively. In the right panel, $\sim10\%$, $\sim1\%$, and $0.04\%$ have deviation larger than 0.1 dex, 0.3 dex, and 1 dex, respectively.

This can be attributed to the different fine energy meshes used for the outer-region exchange calculations. The number of points of the fine energy mesh in F16 is four times larger than that of the present work so that resonances are better resolved in F16. Therefore, we performed another calculation using the same scaling parameters and the same number of points of the fine energy mesh as in F16. The comparison of the effective collision strength between this calculation (data set D) and F16 at $8.82\times10^4~{\rm K}$ is shown in the right panel of Figure~\ref{fig:hb_ecs_062621_mao19_fme16_low}. The difference is negligible even toward the low-temperature end. Since the resonance enhancement is more significant at lower temperatures, the deviation in the right panel of Figure~\ref{fig:hb_ecs_062621_mao19_fme16_high} is smaller than that in the middle panel of Figure~\ref{fig:hb_ecs_062621_mao19_fme16_low}. 

The remaining deviation in the right panel of Figure~\ref{fig:hb_ecs_062621_mao19_fme16_high} is asymmetric ($\Upsilon_2 > \Upsilon_1$). This is attributed to the following two additional causes.

First, some numerical failures are found in the outer-region exchange calculation of F16. Several test calculations are performed, however, we were unable to reproduce the numerical failures. Second, some unresolved resonances (Section~\ref{sct:col}) were found in the outer-region non-exchange calculation of F16. The Perl script used by F16 bypassed the routine to remove the unresolved resonances.   

The above two additional causes explain the remaining deviations in the right panel of Figure~\ref{fig:hb_ecs_062621_mao19_fme16_low}. Since the numerical failures and unresolved resonances are present in the resonance region, effective collision strengths at high temperatures are less affected (cf. the right panels of Figure~\ref{fig:hb_ecs_062621_mao19_fme16_high} and \ref{fig:hb_ecs_062621_mao19_fme16_low}).

\subsection{\ion{S}{XI}}
\label{sct:061611}
\citet[][L11 hereafter]{lia11b} calculated the electron-impact excitation data of \ion{S}{XI} (or ${\rm S^{10+}}$) in a similar approach as the present work. The configuration interaction among 24 configurations was used to calculate the structure (see their Table~1 for more details). The lowest 254 fine-structure levels were included in the close-coupling expansion and the scattering calculation. For simplicity, we limit our comparison to L11 and refer readers to L11 for their comparison with other previous calculations \citet[][$R$-matrix]{bel00} and \citet[][distorted wave]{lan03}.

Both L11 and the present work use the AUTOSTRUCTURE code for the structure calculation. Slightly different scaling parameters are used, yet the energy levels are nearly identical. Energy levels from the present work and L11 agree well with the NIST and MCHF atomic databases, except for the lowest 20 energy levels (the top-right panel of Figure~\ref{fig:sca_lev_clike}). As shown in the top-right panel of Figure~\ref{fig:sca_tran_clike}, transition strengths agree well among NIST, L11, and the present work with merely a few exceptions.

\begin{figure*}
\centering
\includegraphics[width=.6\hsize, trim={4.5cm 1.0cm 1.0cm 1.0cm}, clip, angle=90]{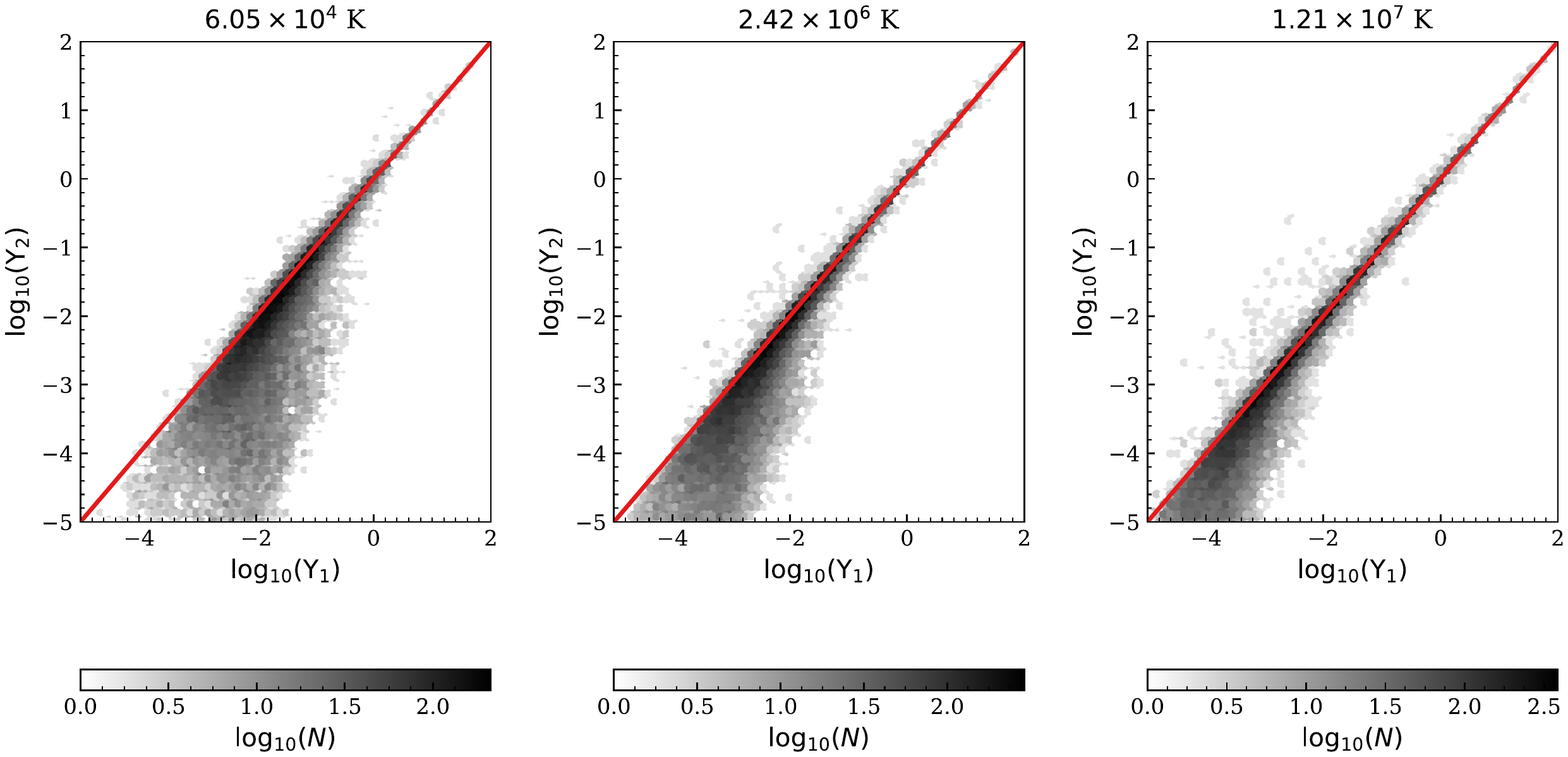}
\caption{Hexbin plots of the comparison of the \ion{S}{XI} (or ${\rm S^{10+}}$) effective collision strengths between the present work ($\Upsilon_1$) and \citet[][$\Upsilon_2$]{lia11b} at $T\sim6.05\times10^4~{\rm K}$ (left) and $2.42\times10^6~{\rm K}$ (middle), and $\sim2.5\times10^5~{\rm K}$ (right). The darker the color, the greater the number of transitions $\log_{10}(N)$. The diagonal line in red indicates $\Upsilon_1=\Upsilon_2$.}
\label{fig:hb_ecs_061611}
\end{figure*}

Both the present work and L11 used the $R$-matrix ICFT method for the collisional calculation. For relatively weaker transitions, the effective collision strengths of L11 are systematically smaller than those of the present work (Figure~\ref{fig:hb_ecs_061611}). There are in total $\sim32000$ transitions with $\log (gf) > -5$ in both data sets, $\sim50-80\%$, $\sim30-50\%$, and $\sim10-20\%$ have deviation larger than 0.1 dex, 0.3 dex, and 1 dex, respectively.

Since the present work has a significantly larger close-coupling expansion (590 levels vs. 254 levels), the additional resonances contribute most to the asymmetric deviation. Similar results were also found by \citet[Fig.~4 of][]{fme16}, where the effective collision strength of \ion{Fe}{XXI} as obtained by two $R$-matrix ICFT calculations with different close-coupling expansions (564 levels vs. 200 levels) were compared. 

\subsection{\ion{Ne}{V}}
\label{sct:061005}
The most recent $R$-matrix calculation of the electron-impact excitation data of \ion{Ne}{V} (or ${\rm Ne^{4+}}$) is presented in \citet[][G00 hereafter]{gri00}. The calculation had 261 fine-structure levels in the configuration-interaction expansion and 138 levels in the close-coupling expansion. Nonetheless, only data for the lowest 49 levels are archived in OPEN-ADAS. 

The energy levels of the present calculation are less accurate (within $\sim10~\%$, the bottom-left panel of Figure~\ref{fig:sca_lev_clike}) compared to NIST and MCHF databases. G00 performed a single configuration MCHF calculation for their atom structure, their energy levels are comparable with the present calculation. The transition strengths also agree well between the present work and G00 (the bottom-left panel of Figure~\ref{fig:sca_tran_clike}). 

\begin{figure*}
\centering
\includegraphics[width=.55\hsize, trim={4.5cm 1.0cm 1.0cm 1.0cm}, clip, angle=90]{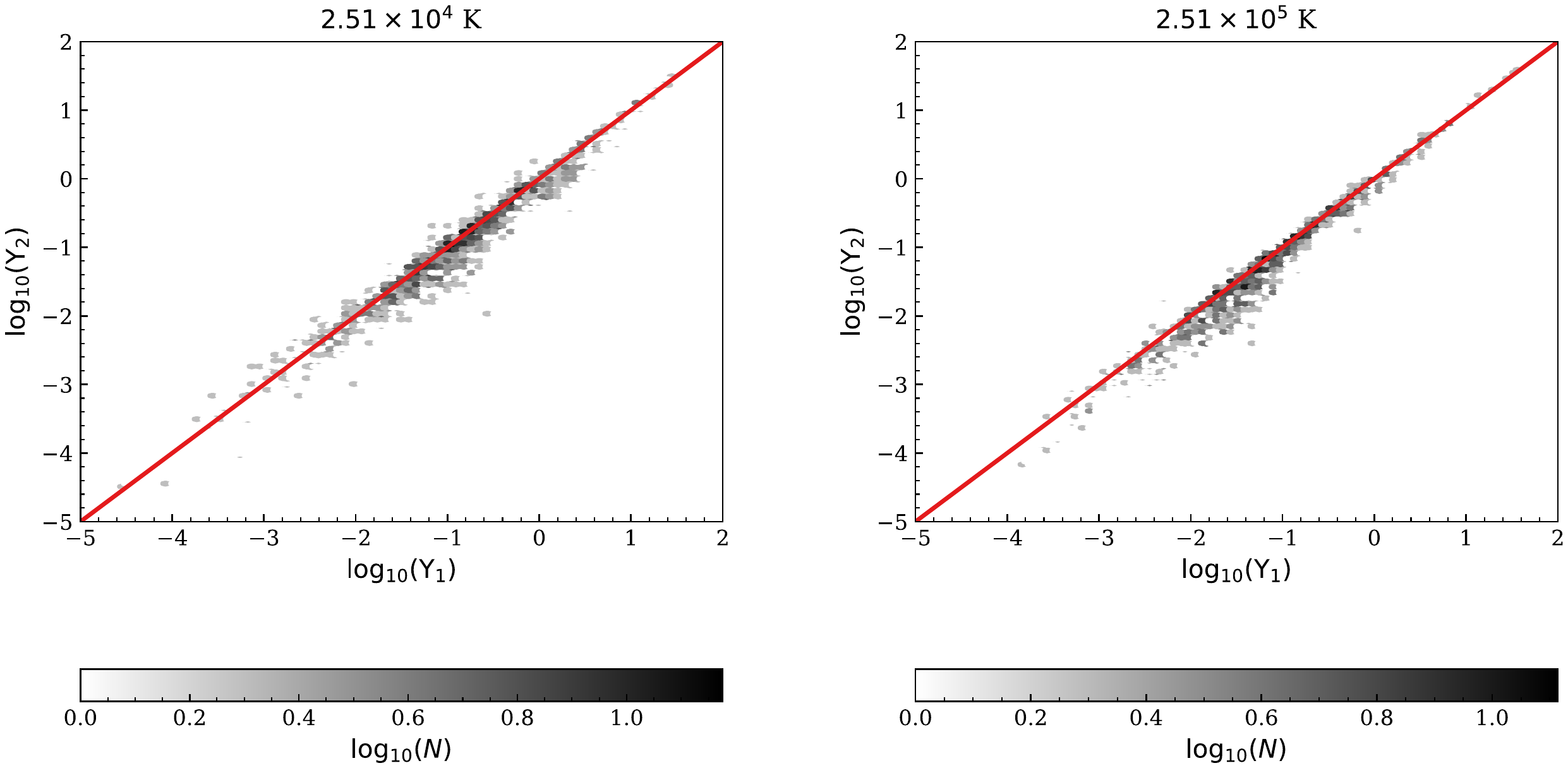}
\caption{Hexbin plots of the comparison of the \ion{Ne}{V} (or ${\rm Ne^{4+}}$) effective collision strengths between the present work ($\Upsilon_1$) and \citet[][OPEN-ADAS, $\Upsilon_2$]{gri00} at $T=2.51\times10^4~{\rm K}$ (left) and $2.51\times10^5~{\rm K}$ (right). The darker the color, the greater the number of transitions $\log_{10}(N)$. The diagonal line in red indicates $\Upsilon_1=\Upsilon_2$.}
\label{fig:hb_ecs_061005}
\end{figure*}

Both G00 and the present work used the $R$-matrix ICFT method for the collision calculation. The effective collision strengths in G00 had a temperature range of $(10^3,~10^6)~{\rm K}$ with three points (1.00, 2.51 and 6.30) per decade. The present calculation covers a different temperature range of $(5\times10^3,~5\times10^7)~{\rm K}$ with three points (1.25, 2.50 and 5.00) per decade. We calculate the effective collision strength of \ion{Ne}{V} at the same temperature points of G00 and show the comparison at at $T=2.51\times10^4~{\rm K}$ and $2.51\times10^5~{\rm K}$ in Figure~\ref{fig:hb_ecs_061005}. We note that the comparison is limited to effective collision strengths involving the lowest 49 levels (all the $n=2$ levels and about a quarter of the $n=3$ levels), which are archived in OPEN-ADAS. There are in total $\sim1170$ transitions with $\log (gf) > -5$ in both data sets, $\sim40-50\%$, $\sim10\%$, and $\sim0.09\%$ have deviation larger than 0.1 dex, 0.3 dex, and 1 dex, respectively. The significant and asymmetric deviation shown in Figure~\ref{fig:hb_ecs_061611} for \ion{S}{xi} is not found here because the results for the low-lying levels are better converged.

\subsection{\ion{O}{III}}
\label{sct:060803}
The most recent $R$-matrix calculation of the electron-impact excitation data of \ion{O}{III} (or ${\rm O^{2+}}$) is presented by \citet[][T17 hereafter]{tay17}, including 202 fine-structure levels in the close-coupling expansion. For simplicity, we limit our comparison to T17 and refer readers to T17 for their comparison with other previous calculations \citep{sto15, pal12, agg99b}. 

T17 used the non-orthogonal MCHF program for their structure calculation, leading to a better agreement of the level energies with respect to the NIST and MCHF atomic databases. As shown in the bottom-right panel of Figure~\ref{fig:sca_lev_clike}, the level energies of the present calculation are less accurate (within $\sim15~\%$). As for the transition strength, strong transitions (i.e., $\log~(gf)\gtrsim10^{-1}$) agree well among all the calculations and databases. A larger deviation is found for some of the weaker transitions (Figure~\ref{fig:sca_tran_clike}). 

\begin{figure*}
\centering
\includegraphics[width=.55\hsize, trim={4.5cm 1.0cm 1.0cm 1.0cm}, clip, angle=90]{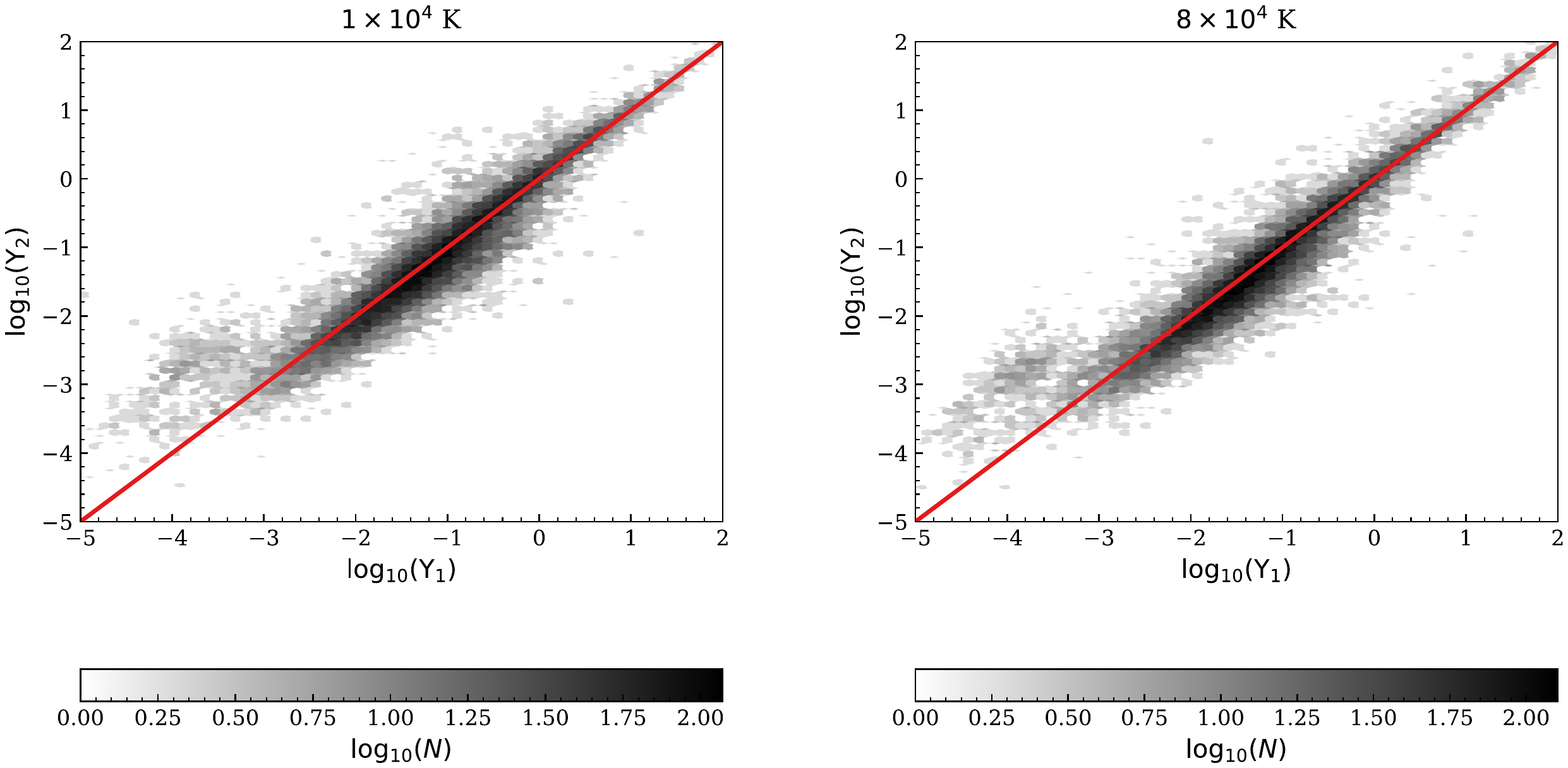}
\caption{Hexbin plots of the comparison of the \ion{O}{III} (or ${\rm O^{2+}}$) effective collision strengths between the present work ($\Upsilon_1$) and \citet{tay17} ($\Upsilon_2$)at $T=10^4~{\rm K}$ (left) and $=8\times10^4~{\rm K}$ (right). The darker the color, the the greater the number of transitions $\log_{10}(N)$. The diagonal line in red indicates $\Upsilon_1=\Upsilon_2$.}
\label{fig:hb_ecs_060803}
\end{figure*}

The scattering calculation of T17 utilized B-spline Breit-Pauli $R$-matrix (BSR) code \citep{zas06}, where an accurate target description is obtained by taking advantage of term-dependent orbitals. The effective collision strengths in T17 are tabulated with a narrower temperature range: $10^{2}~{\rm K}$, $5\times10^{2}~{\rm K}$, $10^{3}~{\rm K}$, $5\times10^{3}~{\rm K}$, $10^{4}~{\rm K}$, $2\times10^{4}~{\rm K}$, $4\times10^{4}~{\rm K}$, $6\times10^{4}~{\rm K}$, $8\times10^{4}~{\rm K}$, and $10^{5}~{\rm K}$. The present calculation covers a wider temperature range of $(1.8\times10^3,~1.8\times10^7)~{\rm K}$ with three points (1.80, 4.50 and 9.00) per decade. We calculate the effective collision strength of \ion{O}{III} at the same temperature points of T17 and show the comparison at $T=10^4~{\rm K}$ and $8\times10^4~{\rm K}$ in Figure~\ref{fig:hb_ecs_060803}. There are in total $\sim19600$ transitions with $\log (gf) > -5$ in both data sets, $\sim60\%$, $\sim20\%$, and $\sim1.5\%$ have deviation larger than 0.1 dex, 0.3 dex, and 1 dex, respectively. The deviation observed is mainly due to the different atomic structures and the different sizes of the close-coupling expansion. The T17 data set is recommended if it suits the purpose of the user.

\subsection{\ion{Ar}{XIII}}
\label{sct:061813}
\begin{figure*}
\centering
\includegraphics[width=.7\hsize, trim={0.cm 0.5cm 1.5cm 0.5cm}, clip]{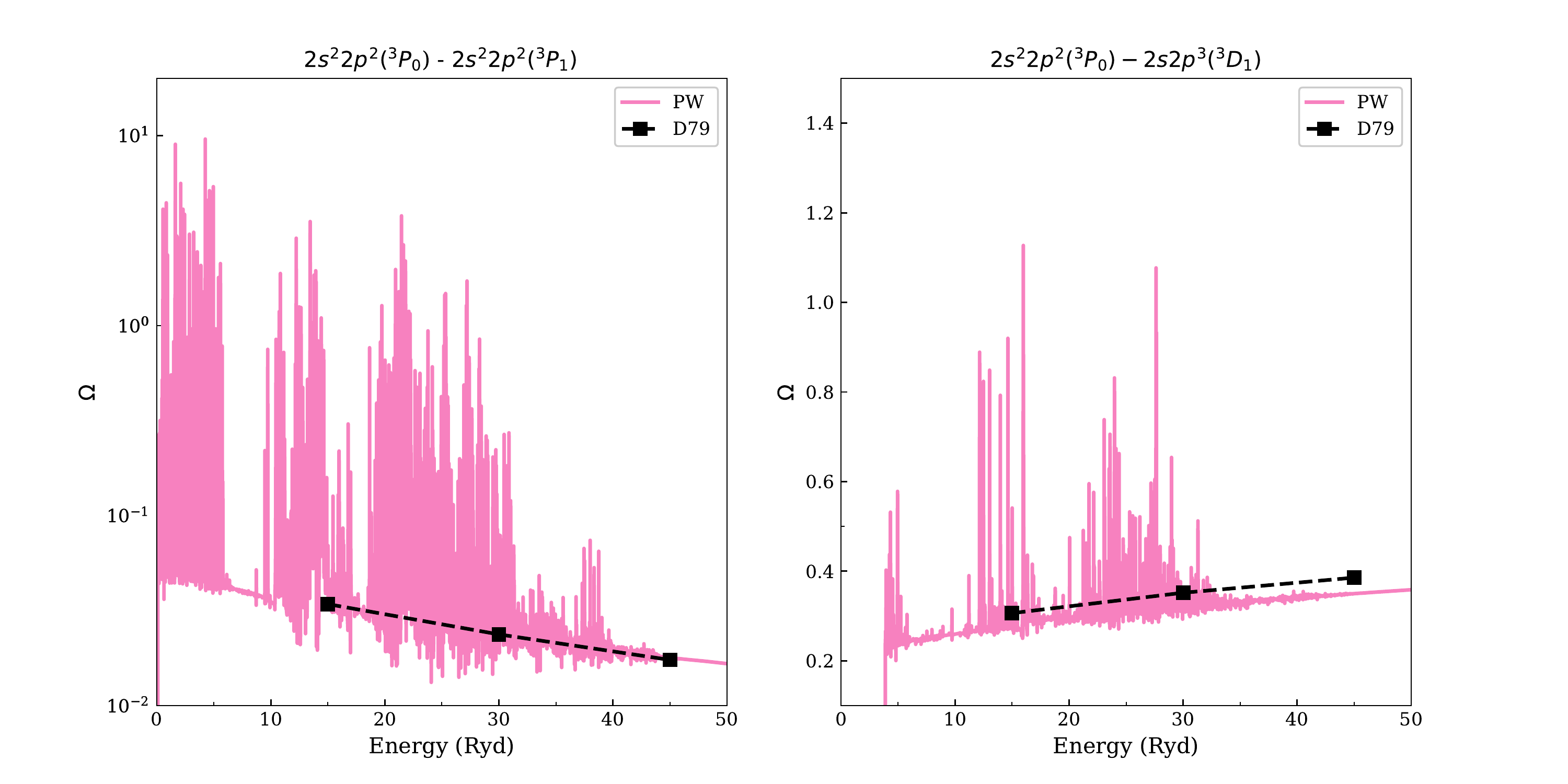}
\caption{Ordinary collision strengths ($\Omega$) for \ion{Ar}{XIII}. The left panel refers to the forbidden transition from the ground level $2s^22p^2~(^3P_0)$ to the metastable level  $2s^22p^2~(^3P_1)$, while the right panel refers to the transition from the ground level $2s^22p^2~(^3P_0) - 2s2p^3~(^3D_1)$. The present work (PW) is shown in pink solid lines. The solid squares and the dashed lines are previous distorted wave approximations from Table 4 of \citet[D79,][]{der79}. }
\label{fig:omega_061813}
\end{figure*}

The collision data of \ion{Ar}{XIII} in the latest version of the CHIANTI database \citep[V9.0][]{der19} originate from \citet{der79}, where the collision calculation was carried out with the UCL distorted wave codes for small angular momentum values of the incoming electron and the Bethe approximation for large angular momentum values. Figure~\ref{fig:omega_061813} compares the ordinary collision strength ($\Omega$) of two transitions from the ground level as calculated with the present $R$-matrix codes and the previous distorted wave calculation. The previous distorted wave calculation provides a good description of the ``background", while the $R$-matrix calculation includes resonances that stand above the background. 

\begin{figure}
\centering
\includegraphics[width=.8\hsize, trim={0.5cm 0.5cm 0.1cm 1.0cm}, clip]{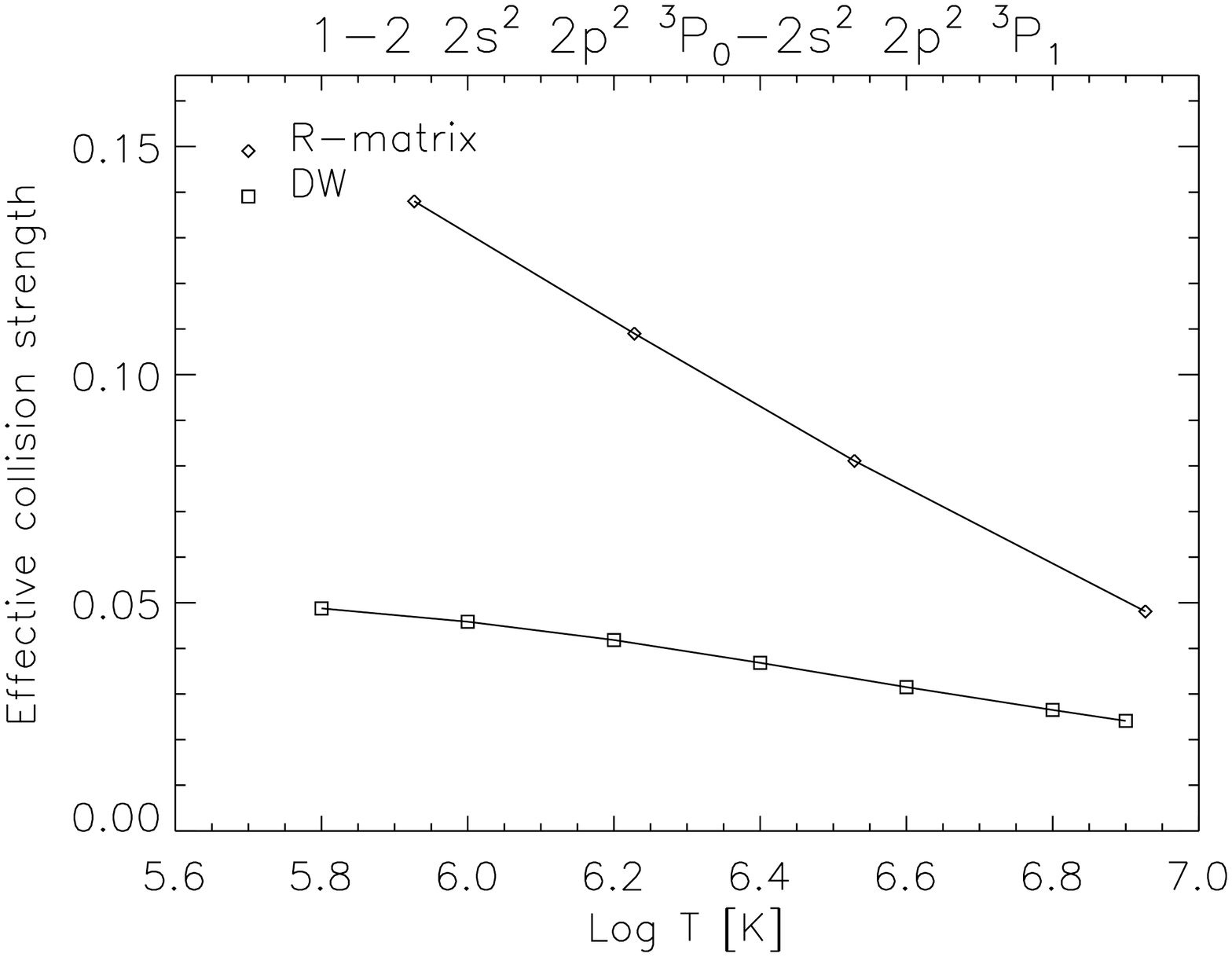}
\includegraphics[width=.8\hsize, trim={0.5cm 0.5cm 0.1cm 1.0cm}, clip]{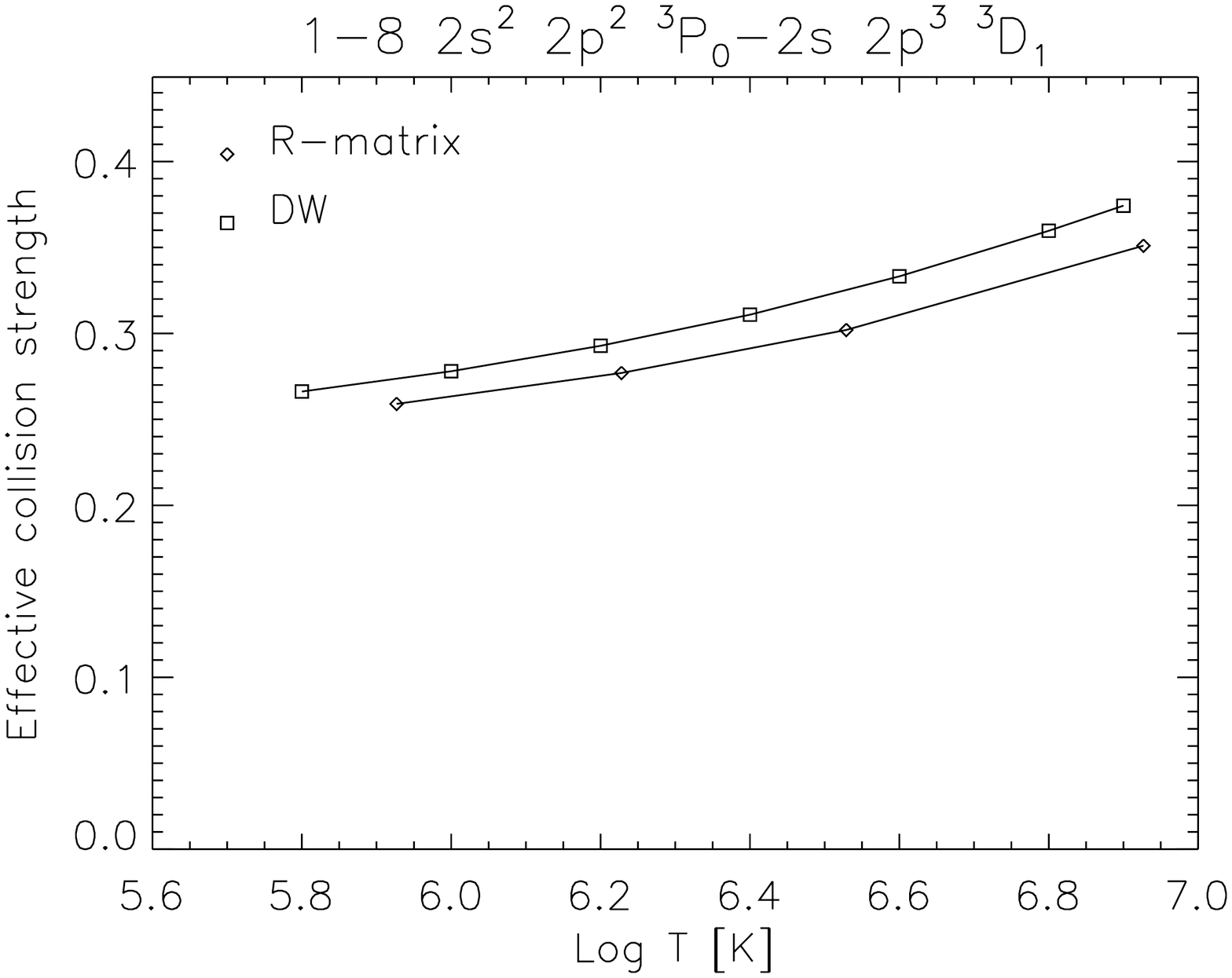}
\caption{Effective collision strengths for one of the strongest forbidden (top) and allowed (bottom) transitions for \ion{Ar}{XIII}. We show both the present values (R-matrix) and those obtained from the CHIANTI database, which were based  on the distorted wave calculations reported by \cite[D79,][]{der79}.}
\label{fig:ar_13_ups}
\end{figure}

It is well known that the presence of the resonances increases significantly the effective collisions strengths for forbidden transitions, especially at low temperatures, while differences for the strong allowed transitions are often small, and mostly dominated by differences in the target structure. Figure~\ref{fig:ar_13_ups} shows the effective collision strengths for the two strongest transitions in \ion{Ar}{XIII}. It confirms that the differences for the allowed transition are minor, while those for the forbidden transition are about a factor of $\gtrsim2$ at the typical formation temperature of this ion in ionization equilibrium ($2.8\times10^6$~K, equivalent to $\log_{10}~(T/K)=6.45$). The distorted wave effective collision strengths were obtained from the CHIANTI database, and are based on the calculations reported by \cite[D79,][]{der79}.

The forbidden line shown in Figure~\ref{fig:ar_13_ups} is one of the two strong infrared transitions which are currently receiving much interest in the solar physics community as they are potentially very useful to measure electron densities and the chemical abundance of Ar, one of the elements for which photospheric abundances are not available. These transitions have never been observed, as they are in a relatively unexplored spectral range; however, they will be observable by the next-generation 4-meter DKIST telescope, as discussed in detail by \cite{dza18b}.

The increased effective collision strengths for all the forbidden transitions we have obtained with the present calculations have a significant effect on their predicted intensities, even though the main populating mechanism for these transitions is cascading from higher levels. To estimate these effects we have considered three ion models. The first has the present effective collision strengths and A-values, but has the A-values for the transitions within the ground configuration from \cite{jon11}. These latter values are obtained with a large-scale GRASP2K calculation and should be very accurate; we note that the differences between these two sets of A-values are small (10--20\%).
The second is the CHIANTI model, but with the A-values within the ground configuration from \cite{jon11}. The third one is the CHIANTI model. 

\begin{figure}
\centering
\includegraphics[width=.9\hsize, trim={0.5cm 0.5cm 0.1cm 0.5cm}, clip]{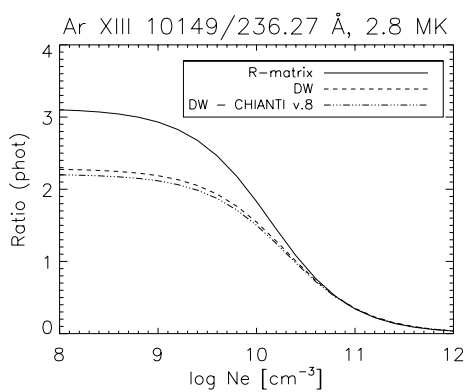}
\caption{The intensity ratio between the main  forbidden and allowed transitions for \ion{Ar}{XIII}, as calculated with the present R-matrix cross-sections and with the 
distorted wave calculations as available in CHIANTI (see text for details).
}
\label{fig:ar_13_ratio}
\end{figure}

Figure~\ref{fig:ar_13_ratio} shows the intensity ratio between the main forbidden and allowed transitions for \ion{Ar}{XIII}, indicating that the cumulative effect of the changes in the effective collision strengths is to increase the intensity of the forbidden line by up to 40\% in the low-density regime.

\section{Summary}
\label{sct:sum}
We have presented a systematic set of $R$-matrix intermediate-coupling frame transfer calculation of C-like ions from \ion{N}{II} to \ion{Kr}{XXXI} (i.e., N$^{+}$ to Kr$^{30+}$) to obtain level-resolved effective collision strengths over a wide temperature range. The present calculation is the first $R$-matrix calculation for many ions in the C-like iso-electronic sequence and an extension/improvement for several ions, with respect to previous $R$-matrix calculations. 

As we have shown for \ion{Ar}{XIII}, the present effective collision strengths increase significantly the predicted intensities of the forbidden lines, compared to earlier calculations. Forbidden lines from \ion{Ar}{XIII}, as well as those from other ions (such as \ion{Si}{IX} and \ion{S}{XI}) are prominent diagnostics for the upcoming DKIST \citep{rim15} solar facility as discussed in \cite{dza18b} and \citet{mad19}. 

The present atomic data will allow more accurate plasma diagnostics with future high-resolution spectral missions such as \textit{Athena} XIFU \citep{nan13,bar18} and \textit{Arcus} \citep{smi16}. For instance, as shown in \citet{kaa17}, \textit{Arcus} has the capability to measure absorption lines from the ground and metastable levels of \ion{Si}{ix}, which enables us to constrain the density of the photoionized outflows in active galactic nuclei.

The effective collision strengths are archived according to the Atomic Data and Analysis Structure (ADAS) data class adf04 and will be available in OPEN-ADAS and our UK-APAP website. These data will be incorporated into plasma codes like CHIANTI \citep{der97,der19} and SPEX \citep{kaa96,kaa18} for diagnostics of astrophysical plasmas. We plan to perform the similar type of calculations for N-like and O-like iso-electronic sequences. 

\begin{acknowledgements}
      The present work is funded by STFC (UK) through the University of Strathclyde UK APAP network grant ST/R000743/1 and the University of Cambridge DAMTP atomic astrophysics group grant ST/P000665/1. JM acknowledges useful discussions with L. Fern{\'a}ndez-Menchero. 
\end{acknowledgements}


\appendix
\section{Comparison of the structure calculation in histograms}
\begin{figure*}
\centering
\includegraphics[width=.7\hsize, trim={1.cm 0.5cm 1.5cm 0.65cm}, clip]{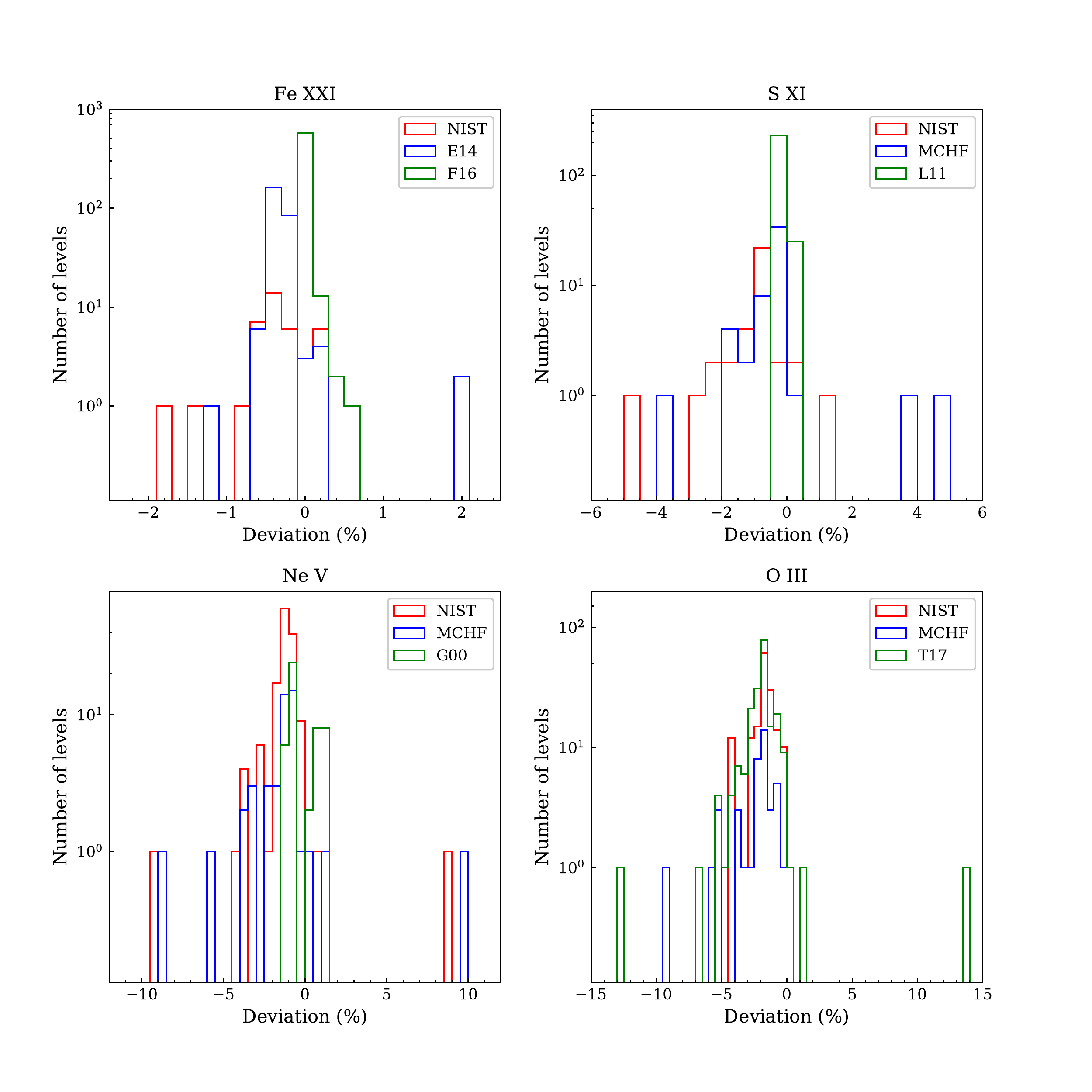}
\caption{Histogram plots of the percentage deviations between the present energy levels (deviation = 0\%), the experimental ones (NIST) and other theoretical values as available in archival databases (MCHF, OPEN-ADAS) and previous works: F16 refers to \citet{fme16}, E14 refers to \citet{ekm14}, L11 refers to \citet{lia11b}, G00 refers to \citet[][OPEN-ADAS]{gri00}, and T17 refers to \citet{tay17}. }
\label{fig:hist_lev_clike}
\end{figure*}

\begin{figure*}
\centering
\includegraphics[width=.7\hsize, angle=90]{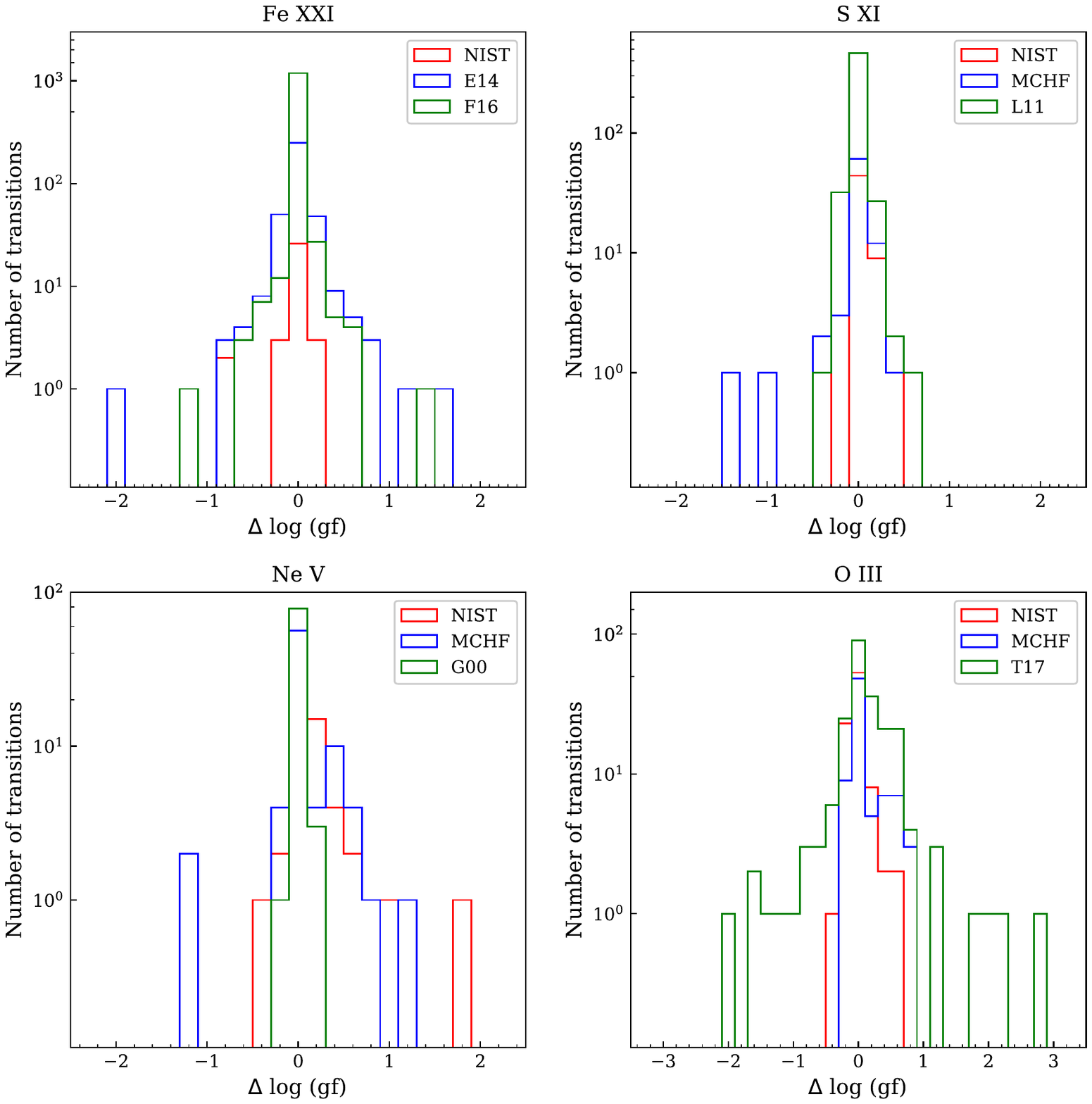}
\caption{Histogram plots of the comparing $\log~(gf)$ in the present work ($\Delta \log~(gf)=0$) with archival databases and previous works. F16 refers to \citet{fme16}, E14 refers to \citet{ekm14}, L11 refers to \citet{lia11b}, G00 refers to \citet[][OPEN-ADAS]{gri00}, and T17 refers to \citet{tay17}. We note that this comparison is limited to relatively strong transitions with $\log~(gf) \gtrsim 10^{-6}$ from the lowest five energy levels.}
\label{fig:hist_tran_clike}
\end{figure*}

We present the histograms (Figure~\ref{fig:hist_lev_clike}) of the energy levels (Figure~\ref{fig:sca_lev_clike}) for the four selected ions: \ion{Fe}{XXI}, \ion{S}{XI}, \ion{Ne}{V}, and \ion{O}{III}. Similar histograms (Figure~\ref{fig:hist_tran_clike}) are presented for $\log~(gf)$ (Figure~\ref{fig:sca_tran_clike}). 

\label{sct:hist_str} 

\section{Energy resolution of the resonance region}
\label{sct:eng_mesh}
\begin{figure*}
\centering
\includegraphics[width=.6\hsize, trim={4.5cm 1.0cm 1.0cm 1.0cm}, clip, angle=90]{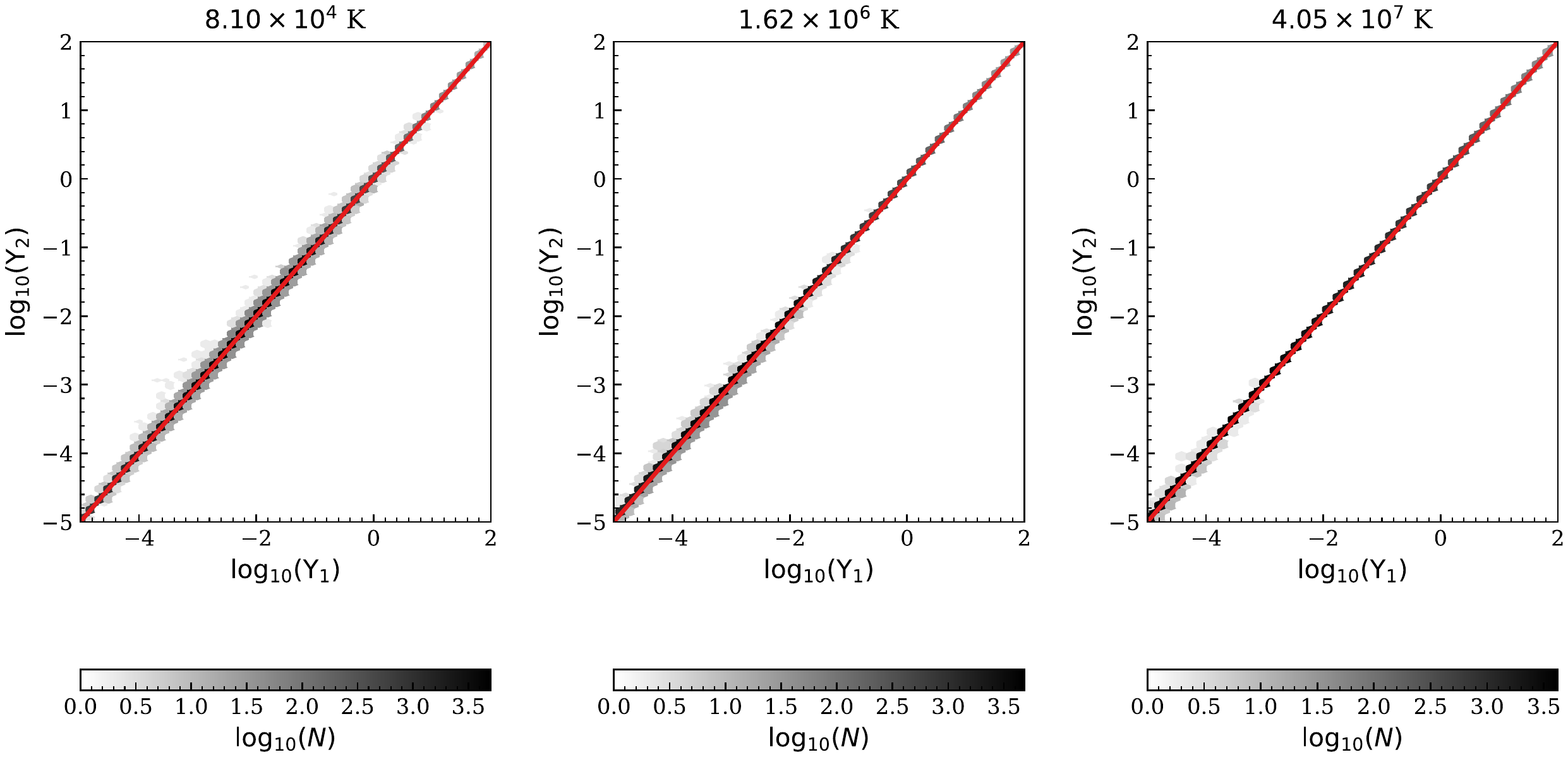}
\caption{Hexbin plots of the comparison of the \ion{Si}{IX} (or ${\rm Si^{8+}}$) effective collision strengths between the default data set of the present work ($\Upsilon_1$) and another calculation where we double the size of the energy mesh for the outer-region exchange calculation ($\Upsilon_2$). The effective collision strengths are compared at $T\sim8\times10^4~{\rm K}$ (left), $\sim1.6\times10^6~{\rm K}$ (middle), and $\sim4\times10^7~{\rm K}$ (right). The darker the color, the greater the number of transitions $\log_{10}(N)$. The diagonal line in red indicates $\Upsilon_1=\Upsilon_2$. }
\label{fig:hb_ecs_061409}
\end{figure*}

For the present calculations of the entire iso-electronic sequence (Section~\ref{sct:col}), our energy resolution of the resonance region is a factor four times larger (poorer) than that of F16 for \ion{Fe}{XXI}. This ``poorer" energy mesh is adequate for low-charge ions like \ion{Si}{IX}. As shown in Figure~\ref{fig:hb_ecs_061409}, we compare the default data set of the present work ($\Upsilon_1$) and another calculation where we double the size of the energy mesh for the outer-region exchange calculation ($\Upsilon_2$). The difference is negligible even toward the low-temperature end. There are in total $\sim1.4\times10^5$ transitions with $\log (gf) > -5$ in both data sets, $\sim0.5\%$, $\sim0.01\%$, and $0\%$ have deviation larger than 0.1 dex, 0.3 dex, and 1 dex, respectively. Thus, the energy mesh used in the present calculation is fine enough for the low-charge ions. 

For high-charge ions like \ion{Fe}{XXI}, at higher temperatures (cf. the middle and right panels of Figure~\ref{fig:hb_ecs_062621_mao19_fme16_high}), the difference between our default data set (A, $\Upsilon_1$) and F16 ($\Upsilon_2$) is mainly due to the (slightly) different atomic structures. At lower temperatures, the scatter caused by different atomic structures is even larger. That is to say, a ``better" atomic structure is the leading concern for high-charge ions at lower temperatures. A ``finer" energy mesh requires more computation time yet only leads to a minor improvement in accuracy. 

\end{document}